\documentclass{article}
\pdfoutput=1
\usepackage{jfrExamplee}
\usepackage{graphicx}
\usepackage{apalike}
\usepackage{setspace}

\usepackage{booktabs} 
\usepackage{enumitem}
\usepackage{cite}
\usepackage{amsmath,amssymb,amsfonts}
\usepackage{algorithmic}
\usepackage{graphicx}
\usepackage{textcomp}
\usepackage{xcolor}
\usepackage{mathtools}
\usepackage{soul}

\usepackage{subcaption}

\newcommand{\cmmnt}[1]{}

\title{Longitudinal Deep Truck\\
{\footnotesize Deep learning and deep reinforcement learning for modeling and control of longitudinal dynamics of heavy duty trucks}
}

\author{
Saleh Albeaik\\
Department of Civil and Environmental Engineering\\
University of California - Berkeley\\
Berkeley, California\\
\texttt{albeaik@berkeley.edu} \\
\And
Trevor Wu \\
Department of Civil and Environmental Engineering\\
University of California - Berkeley\\
Berkeley, California\\
\texttt{trevorhwu@berkeley.edu} \\
\AND
Ganeshnikhil Vurimi \\
Department of Mechanical Engineering\\
University of California - Berkeley\\
Berkeley, California\\
\texttt{gvurimi@berkeley.edu} \\
\And
Xiao-Yun Lu \\
California PATH\\
University of California - Berkeley\\
Richmond, California\\
\texttt{xiaoyun.lu@berkeley.edu} \\
\And
Alexandre M. Bayen \\
Department of Electrical Engineering and Computer Sciences\\
University of California - Berkeley\\
Berkeley, California\\
\texttt{bayen@berkeley.edu}
}

\begin{document}

\maketitle

\begin{abstract}
    Heavy duty truck mechanical configuration is often tailor designed and built for specific truck mission requirements. This renders the precise derivation of analytical dynamical models and controls for these trucks from first principles challenging, tedious, and often requires several theoretical and applied areas of expertise to carry through. This article investigates deep learning and deep reinforcement learning as truck-configuration-agnostic longitudinal modeling and control approaches for heavy duty trucks. The article outlines a process to develop and validate such models and controllers and highlights relevant practical considerations. The process is applied to simulation and real-full size trucks for validation and experimental performance evaluation. The results presented demonstrate applicability of this approach to trucks of multiple configurations; models generated were accurate for control development purposes both in simulation and the field.
\end{abstract}

\section{Introduction}
    Heavy duty truck driving performance is details sensitive \cite{lattemann2004predictive, kirches2013mixed, bae2003parameter, vahidi2003simultaneous, druzhinina2002speed, lu2005heavy}. A typical passenger car could achieve a fairly consistent braking distances while the braking distance of a truck varies significantly within a single trip; before and after hooking up to a trailer, before and after trailer loading, and distance could double as the brakes worms up during the trip. Modeling accuracy has been shown to be a significant (and limiting) factor for precision driving maneuvers \cite{lu2017integrated, spielberg2019neural}. Deep learning has been shown to improve modeling accuracy compared to state of the art classical models for passenger cars \cite{da2019modelling, spielberg2019neural}. However, heavy duty truck modelling literature using deep learning is still sparse. Heavy duty trucks are typically configured and tailor built to optimize to their expected mission requirements. Detailed underlying physics and internal states of trucks are configuration specific and often are different from the more exhaustively modeled variants (components) of passenger cars. In this article, we develop a deep-learning-based longitudinal model for heavy duty trucks and validate its modeling accuracy for heavy duty trucks of different configurations both in simulation and using real-physical trucks. 

\begin{figure*}[!htbp]
        \centering
        \includegraphics[width=\linewidth]{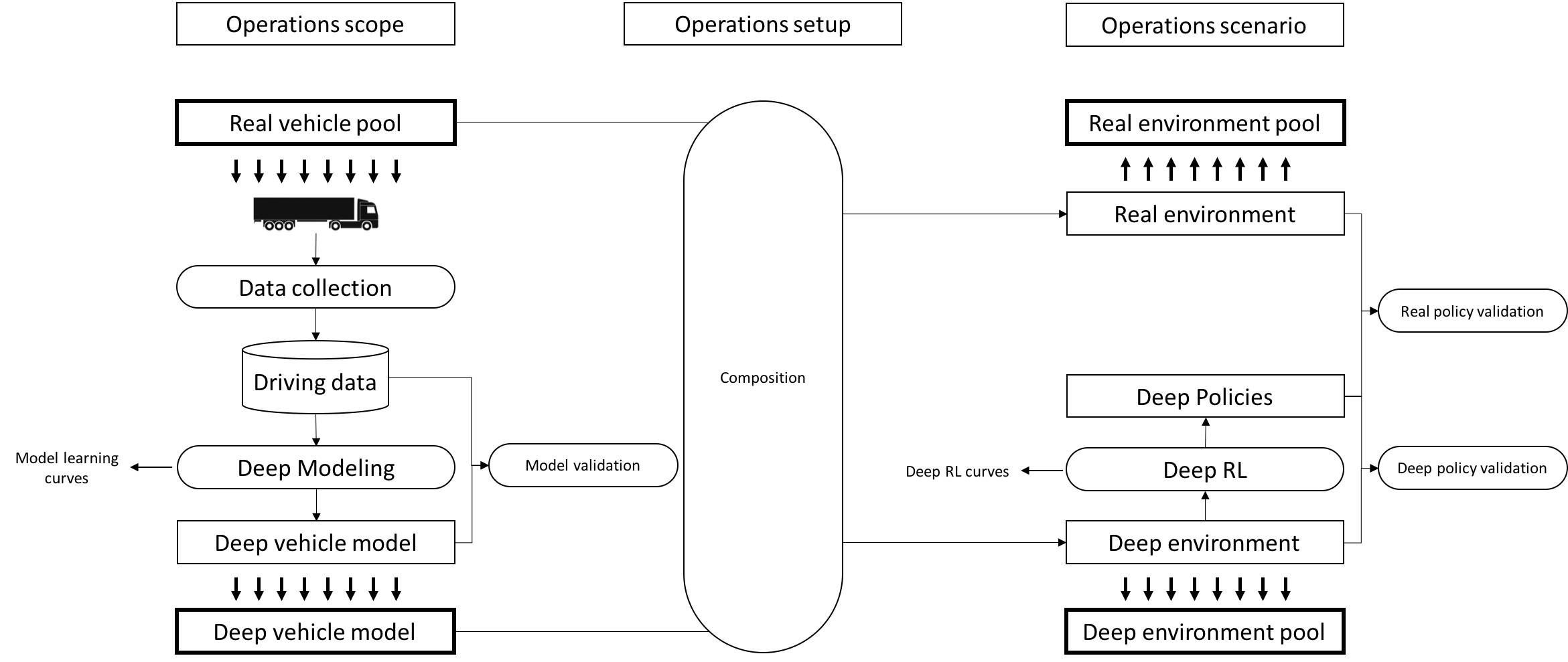} \\
        \vspace{5pt}\hrule\vspace{5pt}
        \includegraphics[width=0.6\linewidth]{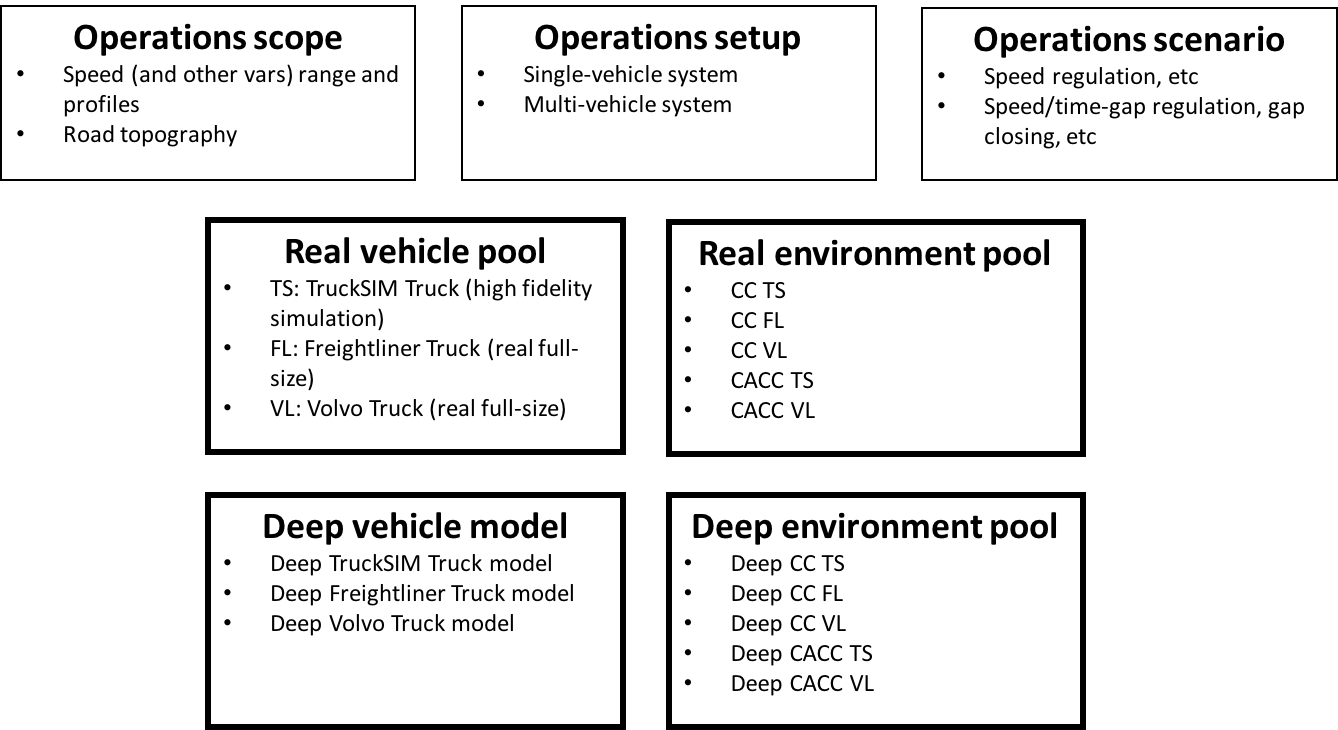}
        \caption{The deep truck process for the development of field-testable deep RL continuous control policies for longitudinal automation of heavy duty trucks and a sample of the pools and operational variations relevant to this work.}
        \label{fig:deep-truck-process-diagram}
    \end{figure*}

    Model-free deep reinforcement learning has been shown to achieve improved performance in many applications in addition to simplifying several previously intractable problems. Transfer of learned policies from simulation is often challenged however by the reality-gap (the mismatch between model and corresponding real-physical system). This article studies the application of deep learning for longitudinal modeling of heavy duty trucks and its application to minimize reality-gap for transferable deep reinforcement learning continuous control policies as shown in Figure~\ref{fig:deep-truck-process-diagram}.
    
    The process uses deep learning to build deep replica models for each truck from some real vehicle pool. These deep replica models are used to develop deep environments suitable for deep reinforcement learning continuous control tasks. The article takes into consideration several of the factors either traditionally or expected to impact modeling and control performance such as vehicle mechanical configuration, operational scope, setup, and traffic scenarios.
    
    Deep learning and deep reinforcement learning offer potential for improved performance at the expense of guarantees such as bounds on control error that are better understood using classical methods. To compensate, more in depth evaluation is always required. To simplify investigation and avoid expulsion of combinatorics however, the article focuses on presenting the process and experimental evaluation of the relevant components and leave additional investigations such as robustness for later articles.

\section{Modeling problem formulation}
\label{sec:deep-learning-model}
    In this article, we formulate heavy duty truck longitudinal dynamics modeling as a time-series supervised deep learning problem. The longitudinal dynamics model $f_{DT}$, detailed in the next section, is represented as:
    \begin{equation}
        {\begin{bmatrix} x(k+1) \\ y(k+1) \end{bmatrix}}
        =
        f_{DT}\left(
            {\begin{bmatrix} u(k) \\ w(k) \end{bmatrix}} 
            \bigg|
            {\begin{bmatrix} x(k) \\ y(k) \end{bmatrix}},
            \Phi
        \right)\label{eq:model-base-form},
    \end{equation}
    where $x$ represent internal state, $y$ represent truck response, $u$ represent controllable inputs to the truck, $w$ represent uncontrollable conditions relevant to the dynamics, and $\Phi$ represent model parameters. The initial conditions are given by $x(k = 0) = x_o$ and $y(k = 0) = y_o$. 
    
    Model parameters $\Phi$ are trained by solving the following optimization problem:
    \[
        \min_{\Phi} \quad \sum_{k}
            {
                \left\lVert
                    \hat{y}(k| \Phi) - y(k)
                \right\rVert_2^2
            }
    \]
    where $\hat{y}$ is the model-based estimate of $y$ given ground truth historical driving time-series data $y$, $u$, $w$, initial state vector $x(k=0)$, and proper recursive substitution of the estimate of the internal state vector $\hat{x}(k| \Phi)$ for $x$ as follows:
    \begin{equation}
        {\begin{bmatrix} \hat{x}(k+1| \Phi) \\ \hat{y}(k+1| \Phi) \end{bmatrix}}
        =
        f_{DT}\left(
            {\begin{bmatrix} u(k) \\ w(k) \end{bmatrix}} 
            \bigg|
            {\begin{bmatrix} \hat{x}(k| \Phi) \\ y(k) \end{bmatrix}},
            \Phi
        \right)\label{eq:model-training-form}.
    \end{equation}
    
    The model is trained using a $K$-step unfolded time-series mini-batch Adagrad (Adaptive stochastic gradient) algorithm \cite{duchi2011adaptive, mcmahan2010adaptive}. Each gradient step is estimated from M independent samples (time-series model evaluations) each of $K$ time steps as follows:
    \[
        \sum_{m = 0 \dots M} \sum_{k = 0 \dots K}
            {
                \left\lVert
                    \hat{y}_{n, m}(k| \Phi) - y_{n, m}(k)
                \right\rVert_2^2
            },
    \]
    where the sub-indices abstract time-series splits and $n = 0 \dots N$ represent the mini-batch index. Training is initialized with random deep network parameters, and with $\hat{x}_{n, m}(k = 0| \Phi) = 0$ for all $n$ and $m$. Given the trained model, truck simulations are generated from:
    \begin{equation}
        {\begin{bmatrix} \hat{x}(k+1| \Phi) \\ \hat{y}(k+1| \Phi) \end{bmatrix}}
        =
        f_{DT}\left(
            {\begin{bmatrix} u(k) \\ w(k) \end{bmatrix}} 
            \bigg|
            {\begin{bmatrix} \hat{x}(k| \Phi) \\ \hat{y}(k) \end{bmatrix}},
            \Phi
        \right)\label{eq:model-deployment-form},
    \end{equation}
    and initialized using $\hat{x}(k = 0| \Phi) = 0$ and $\hat{y}(k = 0) = y_o$, where $y_o$ represent the observable initial condition of truck dynamics. 
    
    Variable instantiations are detailed for each respective experiment in the later sections; however, we assume in general, for longitudinal dynamics, 
    \begin{align*} 
        u(k) &= \begin{bmatrix} E_\text{cmd}(k) \\ B_\text{cmd}(k) \end{bmatrix} \\
        y(k) &= \begin{bmatrix} v(k) \\ a(k) \\ f_\text{rate}(k) \end{bmatrix} \\
        w(k) &= \theta_\text{rdg}(k),
    \end{align*} 
    where $E_\text{cmd}(k)$ is engine command, $B_\text{cmd}(k)$ is brake command, $v(t)$ is vehicle speed, $a(k)$ is vehicle acceleration, $f_\text{rate}(k)$ is fuel rate, and $\theta_\text{rdg}(k)$ is road grade each at discrete time step $k$.
    
    
\section{Deep model}
\label{sec:deep-truck}
    This section presents the \(f_{DT}\) model structure used to represent the longitudinal dynamics of the heavy duty truck. The model assumes that only controllable inputs to the truck, uncontrollable driving environment variables, and truck responses are known and measurable while the configuration of the truck and relevant internal state variables are not specified. 
    
    
    The state model, 
    \[x(k+1) = H(u(k), w(k) | x(k), y(k)),\]
    represents an integrated state observer and tracker equations, and state update and encoder equations. The state model, $H(\cdot)$, is represented in this article as a long short-term memory (LSTM) recurrent neural network (RNN). The use of a single deep network unit to represent this model enables parameter sharing for the state observer, tracker, updater, and encoder functions.
    
    The output model, 
    \[y(k) = G(x(k)),\]
    represents an integrated state decoder equation and explicit output constraints model. The output model is represented by a cascade of a state decoder $D$ and an explicit constraints models $C$ such that $y(k) = C(D(x(k)))$. The state decoder is implemented as a fully connected feedforward neural network parameterized by $\Phi_{nn}$. In the remainder of this article, we implement discrete-time longitudinal kinematics model as an explicit constraint model for longitudinal response output variables:
    \[
        v(k + 1) = v(k) + a(k) \cdot dt,
    \]
    where $v(k)$ and $a(k)$ are longitudinal velocity and acceleration respectively, and $dt$ is discrete time step. Other variables were left unconstrained.
    
    
        
\section{Driving cycles for data collection}
\label{sec:driving-cycles}
    In this article, we assume that the internal dynamics of trucks can be observed from datasets were $y = (v, a)$, $u = (E_\text{cmd}, B_\text{cmd})$, and $w = \theta_\text{rdg}$ are jointly spanning. We consider that, without specialized driving data collection facilities, a human driver is most practical for data collection. Internal truck control signals $u$ are often not accessible though the human driver interface (pedals), however, but are processed though vehicle manufacturer proprietary control systems as shown in Figure~\ref{fig:truck-interfaces-diagram}. We thus approximate such spanning dataset by a driving cycle consisting of (1) $w$ and $y$ spanning arbitrary acceleration/deceleration profiles, (2) $w$ and $v$ spanning coasting ($u = 0$), and (3) $w$ and $B_\text{cmd}$ spanning braking to zero speed.
    
    For the field experiments presented in this article, these instructions were given to a human driver to execute for data collection. In simulation, we used a random generative model to approximate such a driving cycle. The generative model utilizes a moving average random walk model as a road profile generator. Coasting and braking to zero episodes are generated using direct randomized initializations. The spanning arbitrary acceleration/deceleration profiles were generated from the model presented in the next section.

    \subsection{Generative model for state space spanning driving cycles}
        For the numerical experiments presented in this article, we simulate arbitrary driving cycles using a speed profile generative model based on a time adaptive unstable stochastic speed controller. We designed it as a random speed profile (driving cycle) generator that samples the state space ``fairly'' uniformly. 
        
        Double integrator model is used with hard saturation limit at desired maximum and minimum speeds as follows:
        \[
            v(t) = \text{max}(\text{min}(v(t-dt) + a(t) \cdot dt, v_\text{max}), v_\text{min}),
        \]
        where $v_\text{min}$ and $v_\text{max}$ are desired minimum and maximum speeds of generated profile. Acceleration is sampled from a normal distribution as:
        \[
            a(t) = \mathcal{N}(\mu_\text{a, scaling} \cdot \mu_\text{a}(t), \sigma_\text{a, scaling} \cdot \sigma_\text{a}(t)),
        \]
        where $\mu_\text{a, scaling}$ and $\sigma_\text{a, scaling}$ are tuning parameters.
        
        Acceleration statistics are designed based on speed dependant unstable feedback control. Average acceleration is given by:
        \[
            \mu_\text{a}(t) = 1 - \frac{v(T_i)}{v_\text{ref}},
        \]
        where $v_\text{ref}$ is control reference speed, here set to $\frac{v_\text{min}+v_\text{max}}{2}$. The standard deviation is designed to allow for bursts of spontaneous high accelerations but discourage it at extreme speeds ($v_\text{min}$, and $v_\text{max}$) as follows:
        \[
            \sigma_\text{a}(t) = \frac{v(T_i)}{v_\text{ref}} \cdot \left(1 - \frac{v(T_i)}{v_\text{max}}\right).
        \]
        
        $T_i$ is used for acceleration-based adaptive temporal discretization for sampling acceleration statistics, and is designed to make high acceleration episodes short lived. The non-negative integer index is updated to $i \coloneqq i + 1$ and $T_{i+1}$ is re-sampled when $t$ equals $T_{i+1}$ and integrated as follows:
        \[
            T_{i+1} = T_i + \lceil \text{max}(\mathcal{N}(\mu_T) \cdot (1 - | \mu_\text{a}(t) | ), dt) \rceil,
        \]
        where $\mu_T$ is a tuning parameter.
        
        To smooth out the noise, we pass the generated acceleration signal though a moving average filter to get $a_f(t)$, and re-integrate speed with a softmax operator as follows:
        
        \[
            v_f(t) = \frac{\max(v_f(t-1) + a_f(t) \cdot dt, v_\text{min})}{1 + e^{\frac{1}{2} \cdot (\max(v_f(t-1) + a_f(t) \cdot dt, v_\text{min}) - v_\text{max})}},
        \]
        noting that acceleration signal has to be recalculated, as needed, from speed signal after this step.
   
\section{Deep-RL continuous longitudinal control}
\label{sec:deep-rl-control}
    In this article, we use deep-reinforcement-learning to design end-to-end heavy duty truck controllers allowing for offline (1) design/tune the controller, (2) calibrate the controller module to the specific physics of each truck, (3) design an embedded state observer/tracker. In developing these controls we assume limited observable input and output (IO), unknown truck mechanical configuration, and unknown relevant internal state. We formulate the problem as a \emph{Partially Observable Markov Decision Processes} (POMDP) and solve it using deep reinforcement learning framework.
    
    \subsection{POMDP and The deep-RL framework}
    
        Design of continuous control problems can be formulated as a deep-RL problem modeled as a POMDP defined by the tuple \((S, P, OS, OP, A, r, \rho_o, \gamma, T)\), Where \(S\) represents the state space of an RL-environment (system); \(P\) is state transition probability space (governing dynamics of the system); \(OS\) represents the observable state space (space of system outputs); \(OP\) is probability distribution of observation space (governing dynamics of observation model); \(A\) represents the action space (actuation and control variables) of an RL-agent (a decision function, a policy, or a controller); \(r\) is reward function (system performance metric); \(\rho_o\) is initial state distribution; \(\gamma\) is reward discount factor over time; and \(T\) is time horizon. In this manuscript, we use model free Policy Gradient learning \cite{schulman2015trust, duan2016benchmarking} to computationally optimize for expected discounted cumulative reward for an agent policy $\pi_\theta$ parameterized by $\theta$: 
        \[
            \theta^* = \underset{\theta}{\mathrm{argmax}} \sum_{t=1}^{T} {E_{(s_t, a_t) \sim p_{\theta}(s_t, a_t)} \left[ r(s_t, a_t) \right] }
        \]
        where $s_t$ and $a_t$ are state and action at time step $t$, and $p_{\theta}$ is probability distribution over state and action space.
    
    \subsection{Deep-RL cooperative adaptive cruise control}
    \label{sec:deep-rl-cacc}
        \begin{figure}[!htbp]
            \centering
            \includegraphics[width=\linewidth, trim={0cm 0 0.5cm 0},clip]{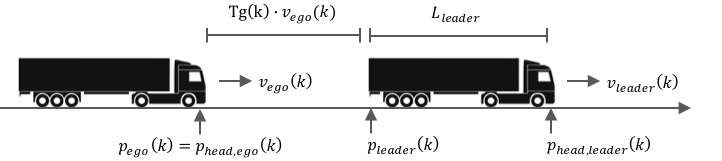}
            \caption{Two truck cooperative adaptive cruise control system setup.}
            \label{fig:two-truck-cacc-illustration}
        \end{figure}
    
        This section formulates an end-to-end two-truck cooperative adaptive cruise control (CACC)~\cite{lu2017integrated} using deep-RL based on the longitudinal truck model developed in this article. In this system, we consider a human driven leader vehicle, and a follower semi-automated to simultaneously regulate speed and time gap as shown in Figure~\ref{fig:two-truck-cacc-illustration}.
    
        The environment is modeled by a two point mass system representing each of the two trucks. Both vehicle dynamics were modeled using the double integrator kinematic model. Leader vehicle (leader) dynamics were simplified as a linear system. The velocity of the controlled truck (ego) is modeled as a nonlinear system according to the deep truck model $f_{DT}$ presented earlier in this article. This results in the following environment model:
        \begin{align*} 
            \begin{bmatrix} 
                p_\text{leader}(k+1) \\
                v_\text{leader}(k+1) \\ 
                p_\text{ego}(k+1) \\
                v_\text{ego}(k+1)
            \end{bmatrix}
            = 
            \begin{bmatrix} 
                1 & dt & 0 & 0 \\
                0 & 1 & 0 & 0 \\ 
                0 & 0 & 1 & dt \\
                0 & 0 & 0 & 0
            \end{bmatrix}
            \begin{bmatrix} 
                p_\text{leader}(k) \\
                v_\text{leader}(k) \\ 
                p_\text{ego}(k) \\
                v_\text{ego}(k)
            \end{bmatrix}
            \\
            +
            \begin{bmatrix} 
                0 \\
                1 \\ 
                0 \\
                0
            \end{bmatrix} \cdot
            a_\text{leader}(k) \cdot dt
            \\
            +
            \begin{bmatrix} 
                0 \\
                0 \\ 
                0 \\
                1
            \end{bmatrix} \cdot
            f_{DT, ego, v}\left(
                {\begin{bmatrix} E_\text{cmd, ego}(k) \\ B_\text{cmd, ego}(k) \\ \theta_\text{rdg, ego}(k) \end{bmatrix}} 
                \bigg|
                {\begin{bmatrix} x_\text{ego}(k| \Phi) \\ v_\text{ego}(k) \end{bmatrix}},
                \Phi
            \right),
        \end{align*} 
        where $f_{DT, \text{ego}, v}$ represents the velocity component from the deep truck model for the ego truck, and relevant variables ($E_\text{cmd, ego}$, $B_\text{cmd, ego}$, $\theta_\text{rdg, ego}$, $x_\text{ego}$, and $\Phi$) are as defined in Section~\ref{sec:deep-learning-model}. As shown in Figure~\ref{fig:two-truck-cacc-illustration}, $p_\text{leader}$, $v_\text{leader}$, and $a_\text{leader}$ are absolute longitudinal position, velocity, and acceleration representing the leading vehicle, and $p_\text{ego}$ and $v_\text{ego}$ are absolute longitudinal position and velocity of the ego vehicle. Time step size is represented by $dt$.
        
        The agent is represented by the probability distribution function $\pi(a_k|o_k, \Phi_\text{agent})$, where $a_k$ represent agent action, $o_k$ represent observation at time step $k$, and $\Phi_\text{agent}$ represent agent parameters. The corresponding control $u(k)$ is implemented as:
        \begin{align*} 
            u(k) &= \begin{bmatrix} E_\text{cmd, ego}(k) \\ B_\text{cmd, ego}(k) \end{bmatrix} = f_\pi\left( \begin{bmatrix} v_\text{leader}(k) \\ v_\text{ego}(k) \\ p_\text{leader}(k) - p_\text{ego}(k) \\ v_\text{ego}(k) \cdot Tg_\text{target} \\ \theta_\text{rdg}(k) \end{bmatrix} \right) \\ &= E\left(\pi\left(a_k \bigg| o_k = \begin{bmatrix} v_\text{leader}(k) \\ v_\text{ego}(k) \\ p_\text{leader}(k) - p_\text{ego}(k) \\ v_\text{ego}(k) \cdot Tg_\text{target} \\ \theta_\text{rdg}(k) \end{bmatrix}, \Phi_\text{agent}\right)\right)
        \end{align*} 
        representing the mean value for a Multi-Layer Perceptron (MLP) Gaussian distribution model.
        
        The reward function is designed to simultaneously regulate time-gap between ego and leader to a given desired time-gap, and regulate velocity of ego to match that the leader. The agent is penalized for actuation cost, here approximated by engine and brake commands. Safety constraint is implemented as a very large penalty term applied when minimum safety distance is violated. The reward function is modeled as:
        \begin{multline*}
            r(k) = - \alpha_p (p_\text{leader}(k)-p_\text{ego}(k)-v_\text{ego}(k) \cdot Tg_\text{target})^2 \\ - \alpha_v (v_\text{leader}(k)-v_\text{ego}(k))^2 - \alpha_E E_\text{cmd}^2(k) - \alpha_B B_\text{cmd}^2(k) \\ - \alpha_\text{crash} \cdot (p_\text{leader}(k)-p_\text{ego}(k) \leq d_\text{safety}),
        \end{multline*}
        where $Tg(k)$ is actual time-gap between leader tail and ego head, $Tg_\text{target}$ is target (desired) time-gap, \(\alpha_x\), \(\alpha_v\), \(\alpha_E\), and \(\alpha_B\) are positive constants, $\alpha_\text{crash}$ is a large positive constant, $d_\text{safety}$ is minimum safety distance, and all other variables are as defined earlier in this section.
        
        
        Each training episode is initialized using leader position $p_\text{leader}(k=0) = 0$, random initial ego truck position error $p_\text{leader}(k=0) - p_\text{ego}(k=0) - v_\text{ego}(k=0) \cdot Tg_\text{target}$ from a \(\text{uniform} (p_{o, min}, p_{o, max})\), random initial leader speed \(v_\text{leader}(k=0)\) from \(\text{uniform} (v_{o, min}, v_{o, max})\), random initial ego truck speed error \(v_\text{ego}(k=0) - v_\text{leader}(k=0)\) from \(\text{uniform} (v_{o, min}, v_{o, max})\) distribution, random desired time gap $Tg_\text{target}$ from a \(\text{uniform} (Tg_{o, min}, Tg_{o, max})\) distribution, and random constant road grade $\theta_\text{rdg}$ from a \(\text{uniform} (\theta_\text{rdg, o, min}, \theta_\text{rdg, o, max})\) distribution. To simplify the setup, we also assume $a_\text{leader}(k) = 0$. All distribution boundaries are positive constants chosen to cover the desired operational state-space of the CACC system and be constrained by the state-space covered by the deep model where appropriate.
        
\section{Vehicle pool}
\begin{figure*}[!htbp]
        \centering
        \includegraphics[width=\linewidth]{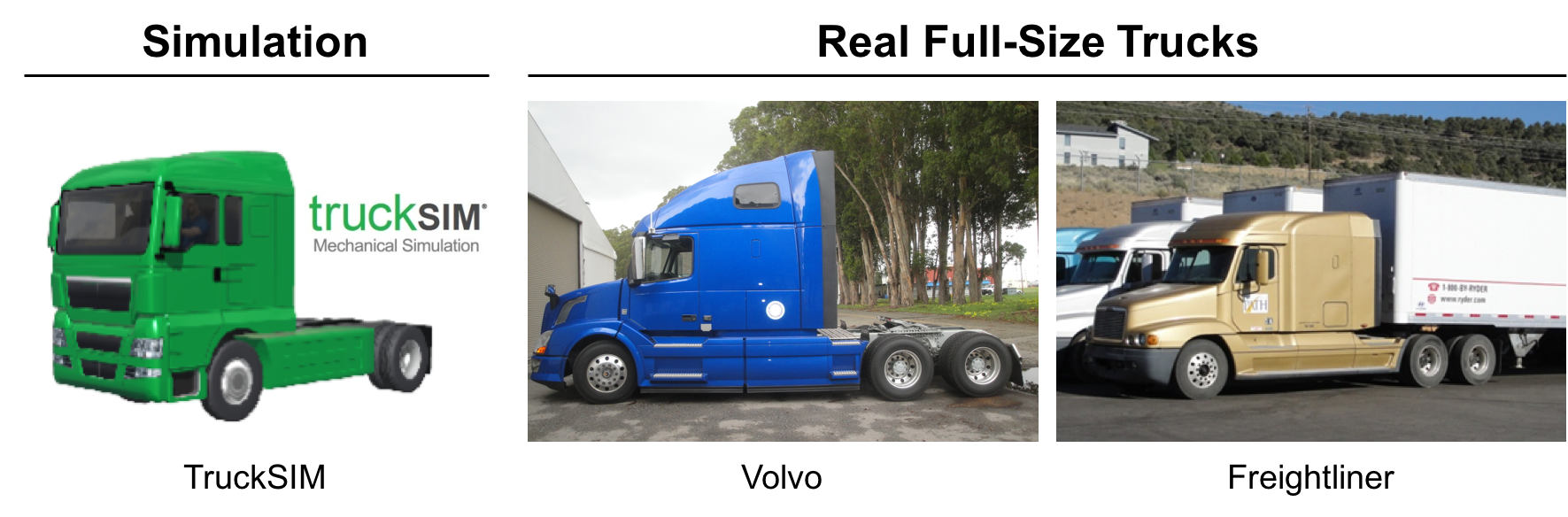}
        \caption{Real full-size and simulations trucks of multiple mechanical configurations used in this research.}
        \label{fig:truck-pool-images}
    \end{figure*}

    We primarily utilize three trucks with three different mechanical configurations for the study presented in this article as shown in shown in Figure~\ref{fig:deep-truck-process-diagram} and Figure~\ref{fig:truck-pool-images}. One truck is simulation based used primarily for numerical experiments. The remaining two trucks are full-size real-physical trucks that had been modeled using two different physics-based power-train models in \cite{XYLu2005HDVModelandLongControl} and \cite{lu2017integrated} and used to develop high precision control systems within each respective article.
    
    \textbf{Simulation framework and simulation truck mechanical configuration.} Simulation experiments in this article are conducted in TruckSim \cite{sayers1996modeling}, a black-box state-of-the-art commercial software framework with high fidelity modeling capabilities and a detailed vehicle and vehicle component libraries. 
    
    The truck, shown in Figure~\ref{fig:truck-pool-images}, is equipped with a 402hp engine. The engine shaft is connected to one side of the transmission via clutch. The clutch allows speed difference between the engine and the transmission when gear shifts. The transmission has ten forward gears and one reverse gear. The other side of the transmission is connected to rear wheels via a differential gear with a fixed reduction ratio. The truck is equipped with an air-brake system. The front air-brakes have capacity of 7.5 kN-m on each wheel. The rear brakes have capacity of 10 kN-m on each wheel. Actuation control input to the truck are engine torque and brake cylinder pressure. The details are presented here for completeness and for reporting purposes, but are irrelevant to the deep model. 
        
    \textbf{Real full-size Freightliner truck mechanical configuration. } The Freightliner truck used for the results in this section is a tractor-only Freightliner Century truck driven by a 435 hp turbocharged Detroit Diesel diesel engine and equipped with a 6 gear true-automatic (equipped with torque-converter) Allison transmission system. The service brake is a drive by wire all the way to the wheels. The truck is not equipped with road grade sensors. 
            
    \textbf{Real full-size Volvo truck mechanical configuration. } The second set of experiments were conducted using a Volvo VNL truck (with and without a tractor) driven by a 500 hp engine. The mechanically most significant differentiator of this truck from the Freightliner truck is the transmission system which is an automated manual-transmission (equipped with clutches).
       
\section{Vehicle interface}
\begin{figure*}[!htbp]
        \centering
        \includegraphics[width=\linewidth]{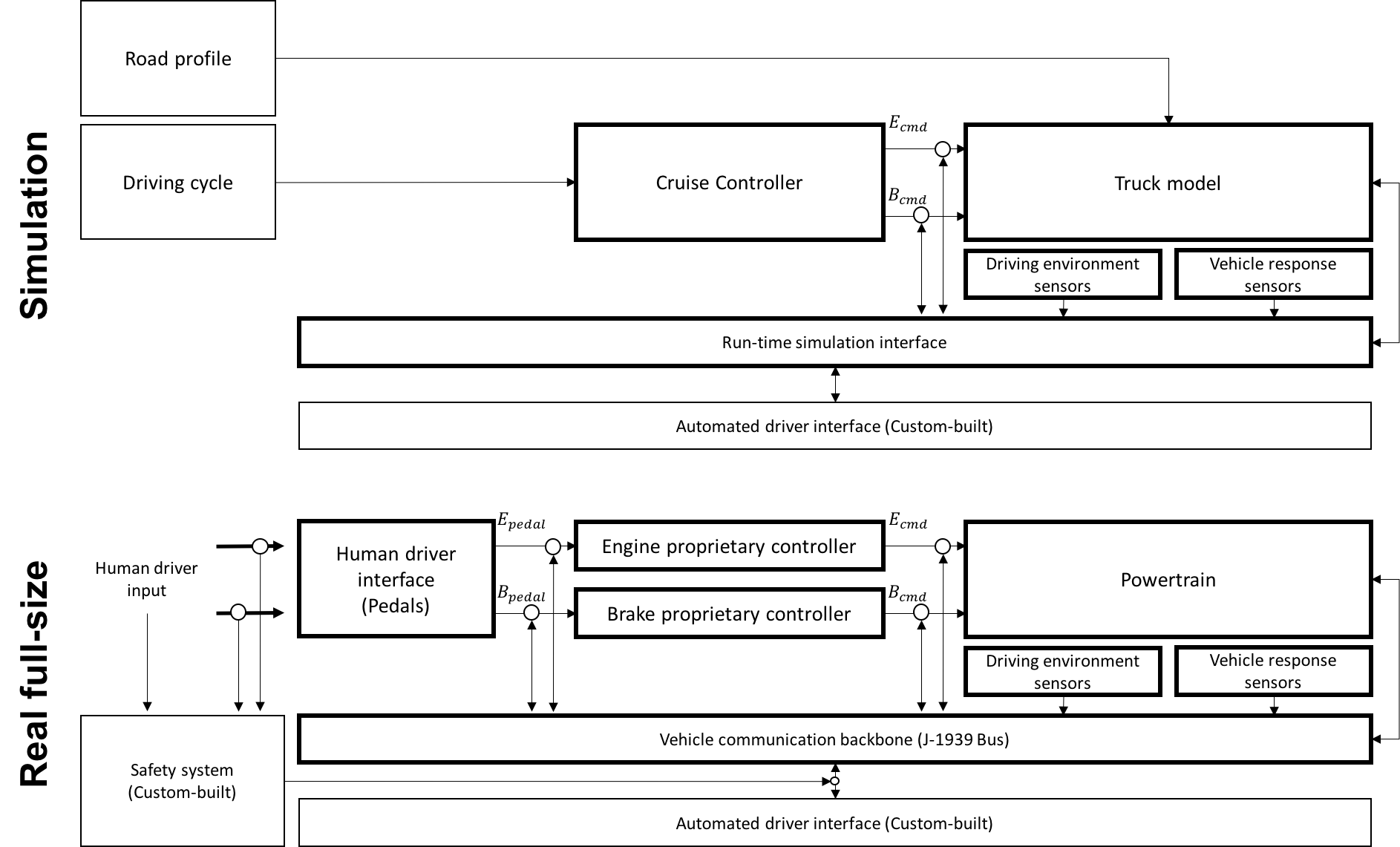}
        \caption{Interface architecture for deep modeling and control of heavy duty trucks.}
        \label{fig:truck-interfaces-diagram}
    \end{figure*}

    Access to vehicle powertrain is often primarily provided through a human driver interface (pedals) and is mediated by proprietary controllers as shown in Figure~\ref{fig:truck-interfaces-diagram}. For precision sensitive applications however, it is often desirable to probe as close to the powertrain as possible (e.g. engine torque or engine fuel rate control signals). We access these signals through a custom-built automated driver interface connected to vehicle communication backbone J-1939. The interface provide access to powertrain and sensor signals; however, architectural details and signal accessibility vary between truck platforms. Multiple layers of fail-safe safety systems were implemented to ensure experiments remain faithful to published description while maintaining safety on the road. Parallel interfaces and system architecture is used for the simulation truck.
    
\section{Experiments}
    This section presents experimental evaluation of the process detailed in this article. The section starts by applying the process to a simulation based truck to present detailed performance statistics. The section then reapplies the process to full-size trucks.

    \subsection{Deep modeling of the simulation truck}
        This section presents experimental results for the development of a deep learning model as described in Section~\ref{sec:deep-learning-model} for the simulation truck.

        \subsubsection{Deep model specifications}
            In this experiment, the uncontrollable conditions $w(k) = \theta_\text{rdg}(k) [\%]$ represent road grade. The controllable input to the truck is given by $u(k) = [ E_\text{cmd}(k), B_\text{cmd}(k)], $ where $E_\text{cmd}(k)$ is engine torque in [$N-m$] and $B_\text{cmd}(k)$ is service brake master cylinder pressure [$0-100\%$].
                
            The output (truck response) vector is given by $y(k) = [a(k), v(k), F_\text{rate}(k)],$ where $a(k)$ is longitudinal acceleration in [$m/s^2$], $v(k)$ is longitudinal speed in [$m/s$], and $F_\text{rate}(k)$ is fuel rate in [$cm^3/s$].
            
        
        \subsubsection{Driving datasets}
\begin{figure}[!htbp]
                \centering
                \includegraphics[width=0.85\linewidth]{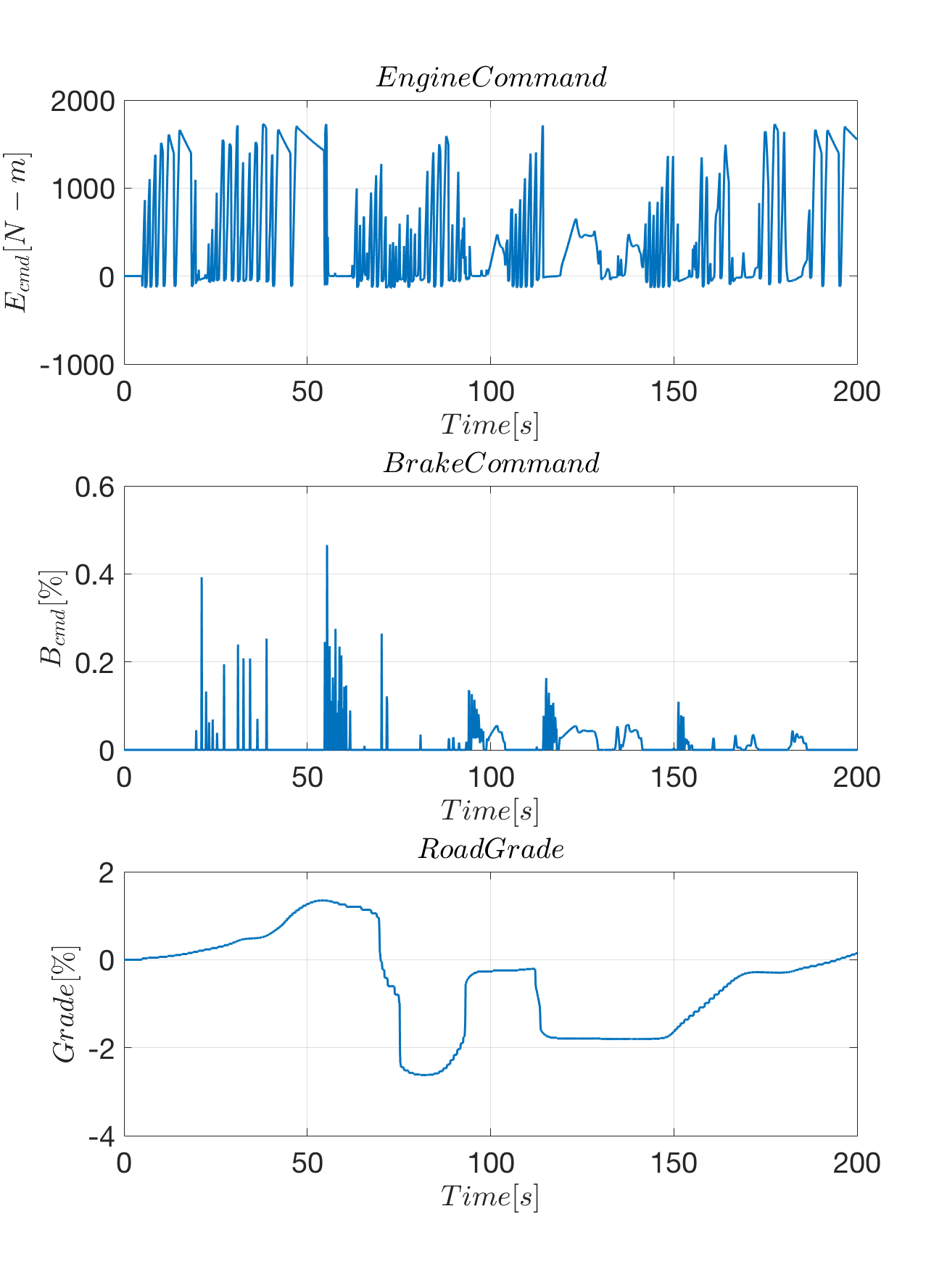}
                \caption{A ground truth sample dataset representing inputs to the deep model.}
                \label{fig:trucksim-sample-dataset-model-inputs}
            \end{figure}

\begin{figure}[!htbp]
                \centering
                \includegraphics[width=0.85\linewidth]{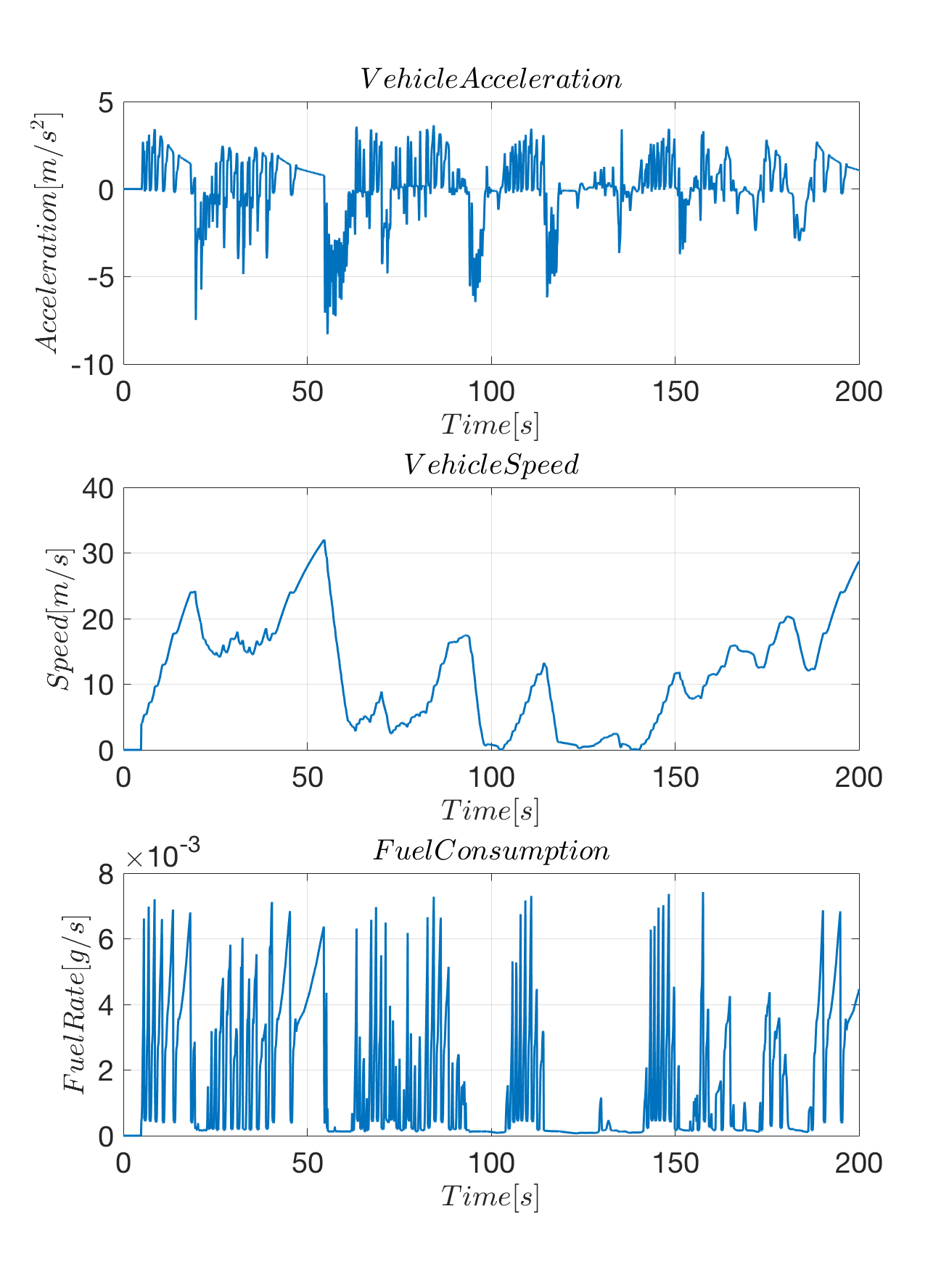}
                \caption{A ground truth sample dataset representing outputs from the deep model.}
                \label{fig:trucksim-sample-dataset-model-outputs}
            \end{figure}

            For training, we simulated a total of four hours of driving using the data collection strategy presented in Section~\ref{sec:driving-cycles}. We generated another three hour set for testing and validation to evaluate modeling performance on unseen data. All datasets span speeds from zero to $35$ $m/s$ and road grades from $\pm3\%$. A sample of the dataset is shown in Figure~\ref{fig:trucksim-sample-dataset-model-inputs} and Figure~\ref{fig:trucksim-sample-dataset-model-outputs}.
            
        \subsubsection{Model learning curves}
            \begin{figure}[!htbp]
                \centering
                \includegraphics[width=\linewidth]{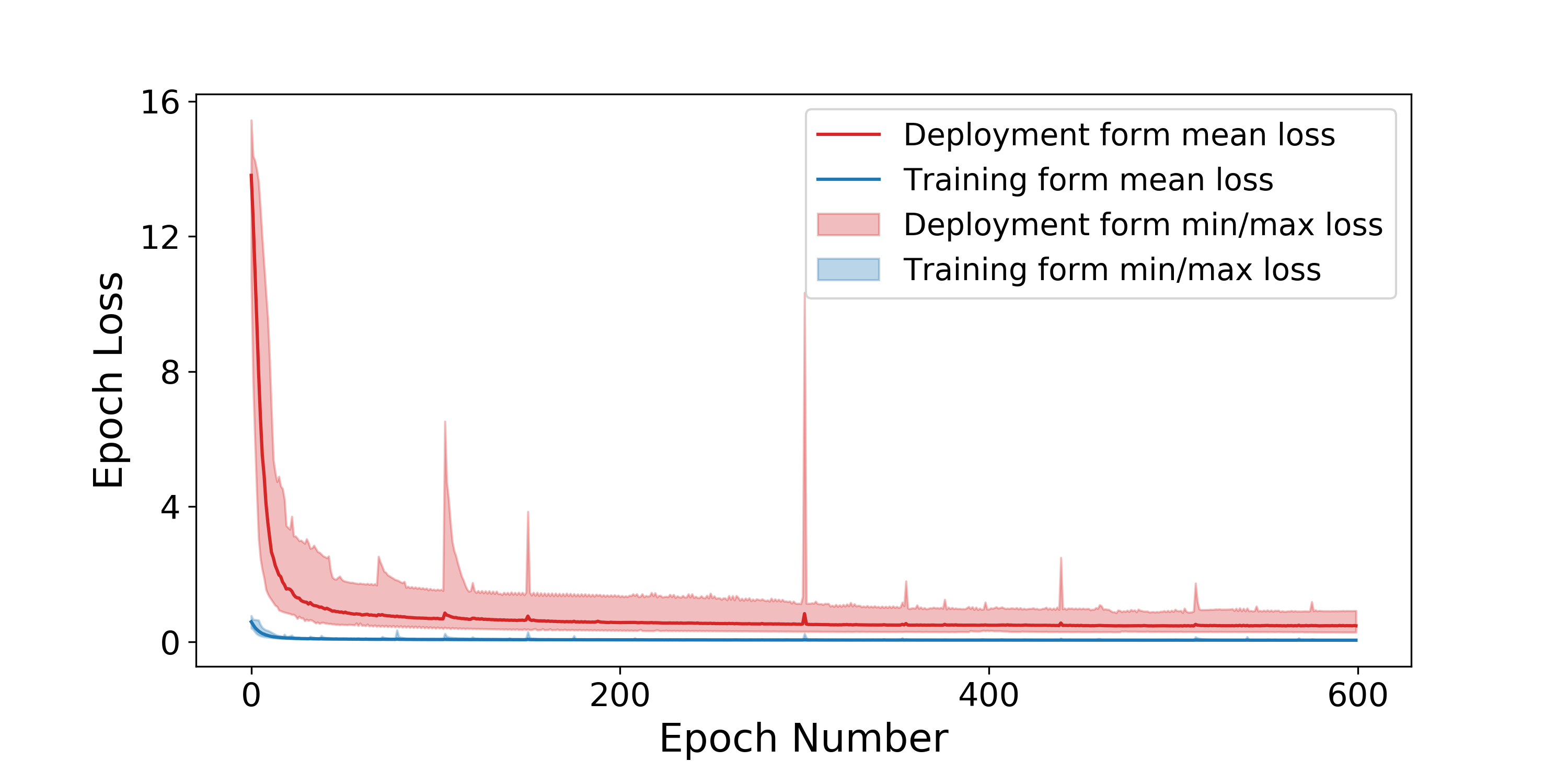}
                \caption{Learning curve---min/max and mean from 30 seeds---for deep modeling of the TruckSIM truck.}
                \label{fig:trucksim-model-learning-curve}
            \end{figure}
            
            This section presents learning curves for training the deep truck model for the simulation truck using the data presented in this section. Figure~\ref{fig:trucksim-model-learning-curve} shows the loss function statistics (mean, min and max) based on training form Equation~\eqref{eq:model-training-form} and deployment form Equation~\eqref{eq:model-deployment-form} from 30 seeds. Each curve is produced using a separate dataset both unseen during training. 
            
            Both learning curves stabilize and converge by the 600th epoch. They exhibits spikes we speculate are a symptom of the inherent stochasticity of the mini-batch algorithm we used. An expected loss gap between training form curve and deployment form curve is observed. 
        
        \subsubsection{Results and model validation}
\begin{figure}[!htbp]
                \centering
                \includegraphics[width=0.8\linewidth, trim={18cm 0 16cm 0},clip]{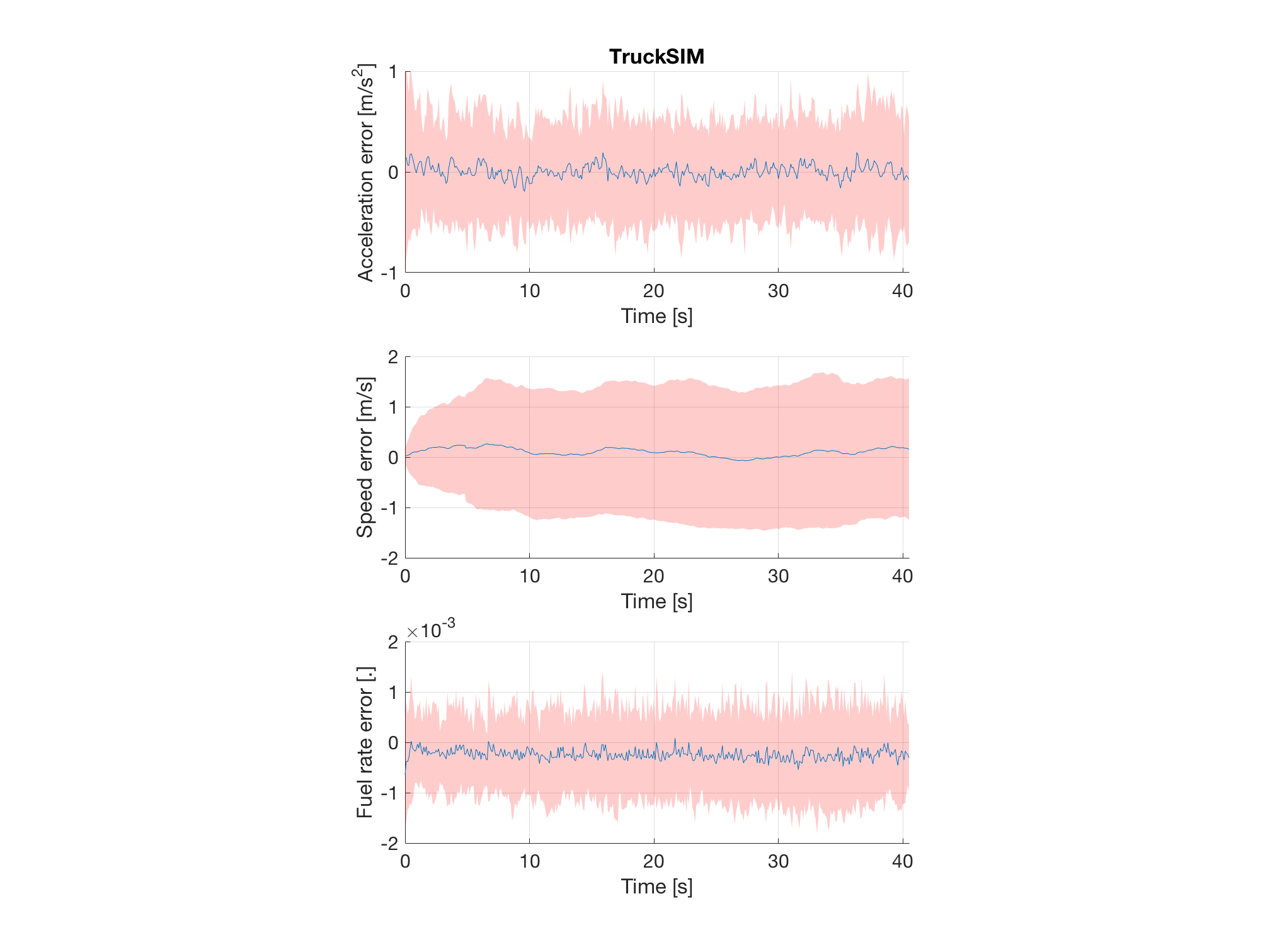}
                \caption{Model error statistics---standard deviation (red shaded areas) and mean (blue curves) from 90 trials---for deep modeling of the TruckSIM truck.} 
                \label{fig:sim-modeling-error-stats}
            \end{figure}

\begin{figure}[!htbp]
                \centering
                \includegraphics[width=0.70\linewidth, trim={0cm 4cm 0cm 2cm},clip]{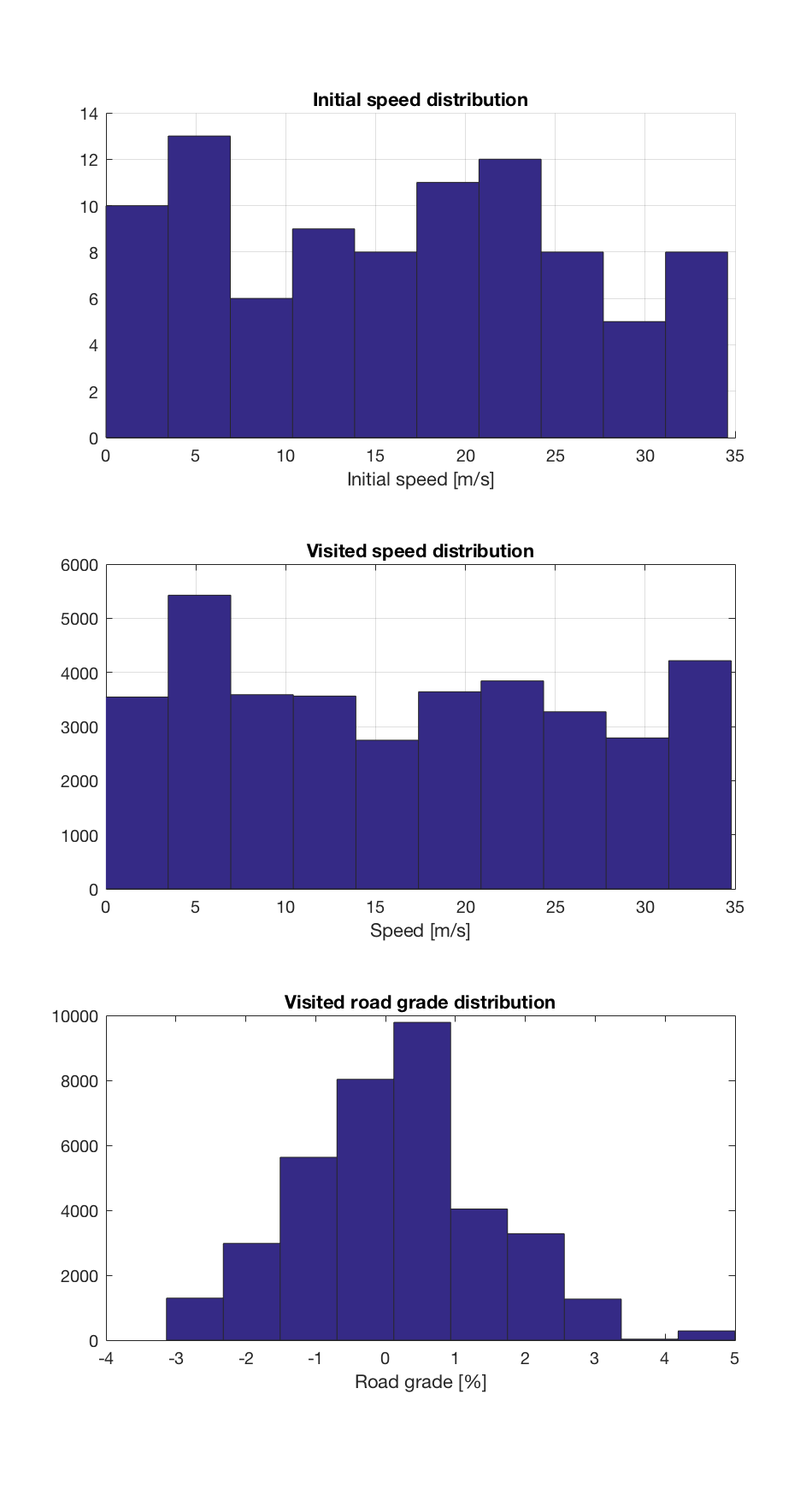}
                \caption{Scenario distribution of the dataset used to compute model validation statistics presented in Figure~\ref{fig:sim-modeling-error-stats}.} 
                \label{fig:sim-modeling-validation-dataset-distributions}
            \end{figure}

            This section presents model validation results using an unseen validation dataset. Figure~\ref{fig:sim-modeling-error-stats} shows modeling error statistics as a function of model simulation time from 90 independent random trails. Error statistics are generated as:
            \[\text{ErrorStatistic}(k)= \text{Statistic}_m\left(\hat{y}_{m}(k| \Phi) - y_{m}(k)\right)\]
            where $m$ is trial number, and $\hat{y}$ follow the deployment form Equation~\eqref{eq:model-deployment-form}. Distributions (initial speed distribution, visited speed over time and across all trials, visited road grades over time and across all trials) of the validation dataset are shown in Figure~\ref{fig:sim-modeling-validation-dataset-distributions}.
            
            Mean of modeling error stays bounded near zero over the 40 second simulation time. On average acceleration deviates by less than $0.5 m/s^{2}$ and fuel deviates by less than $10^{-3}$ at any given time. The statistics also show that the error of modeled speed is expected to remain within $1.5m/s$ over a 40 second simulation time.
            
\begin{figure}[!htbp]
                \centering
                \includegraphics[width=0.85\linewidth]{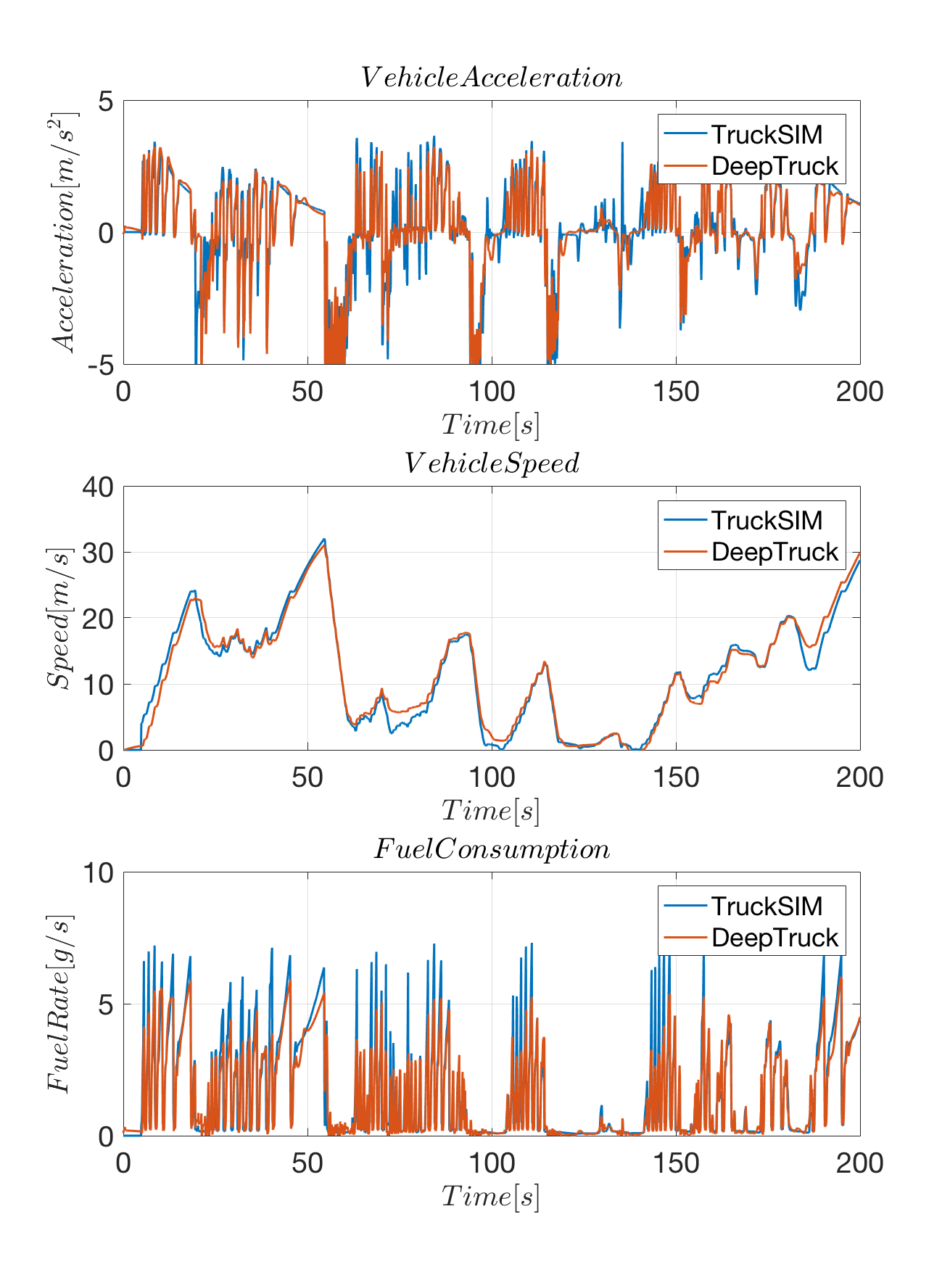}
                \caption{A sample ground truth and model output using a sample validation set.}
                \label{fig:trucksim-model-validation-sample-dataset}
            \end{figure}

\begin{figure}[!htbp]
                \centering
                \includegraphics[width=0.85\linewidth]{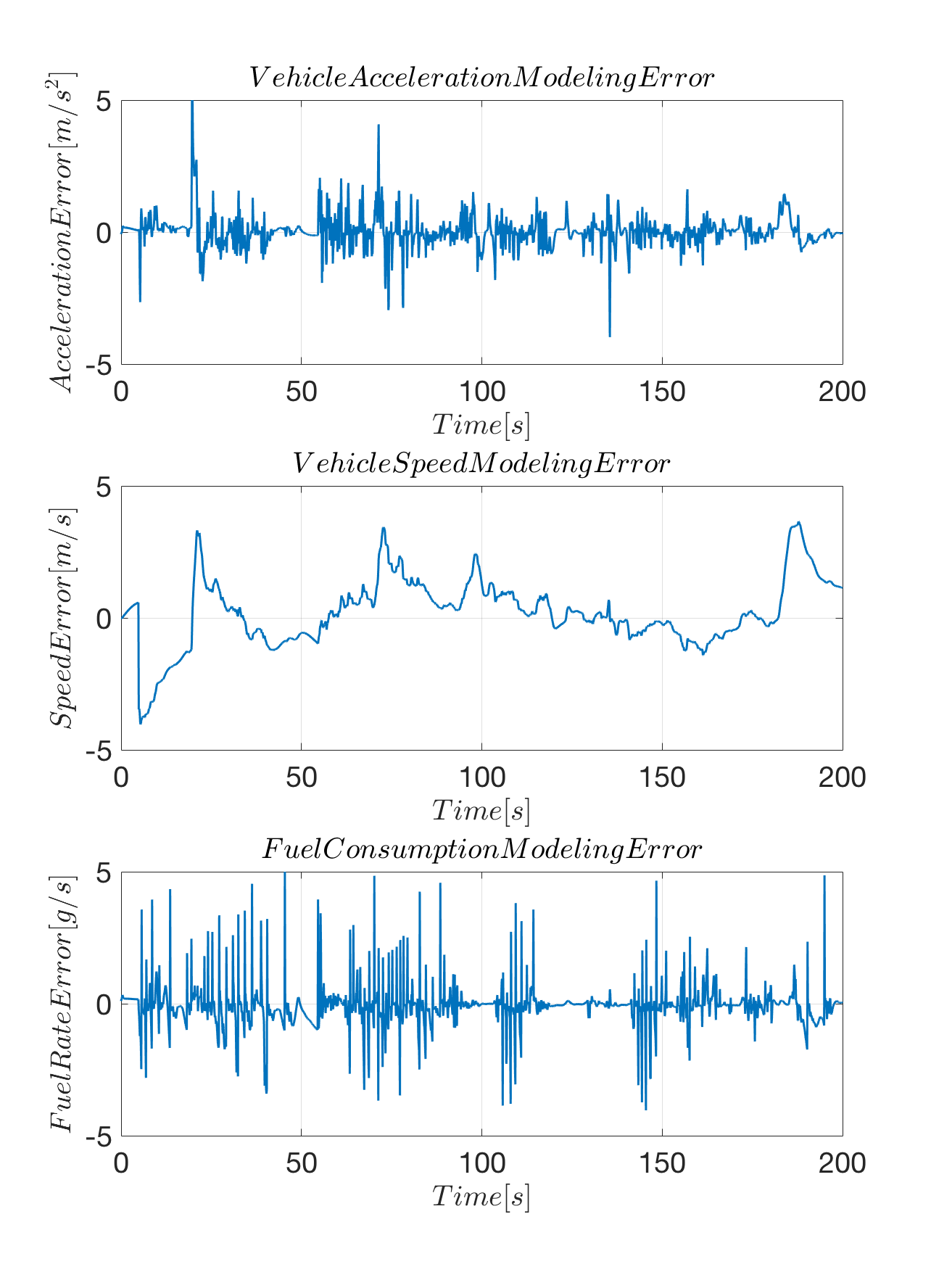}
                \caption{A sample error over time between ground truth and model output using a sample validation set.}
                \label{fig:trucksim-model-validation-error-sample-dataset}
            \end{figure}

            A sample model validation dataset is shown in Figure~\ref{fig:trucksim-model-validation-sample-dataset} and Figure~\ref{fig:trucksim-model-validation-error-sample-dataset}. In this validation experiment, the model is initialized once at $k = 0$ and then simulated for 2000 time steps ($t_\text{end} = 200s$). The dataset exhibits a large initial error transient with significant model response delay estimation error. Error statistics appear to be (by visual inspection) stationary consistent with error statistics in Figure~\ref{fig:sim-modeling-error-stats}.
            
        
    \subsection{Deep-RL control of the simulation truck}
        
        This section presents experimental results for the development of a deep-RL CACC as described in Section~\ref{sec:deep-rl-control} for the simulation truck. 
            
        \subsubsection{Training setup and learning curves}
            For this experiment, the sampling rate is set to $10Hz$ ($dt = 0.1s$) and we assume flat driving environment with no other relevant driving environment variables; thus we substitute $w(k)$ with the empty set. The controllable input to the ego truck (agent output) is given by $u(k) = [E_\text{cmd, ego}(k), B_\text{cmd, ego}(k)],$
            where $E_\text{cmd, ego}(k)$ is requested engine torque in [$N-m$] and $B_\text{cmd, ego}(k)$ is requested service brake master cylinder pressure percentage [$0-100\%$]. The agent $\pi$ is modeled using an ANN that has 3 hidden layers, each of size 25.
            
            Each training episode is initialized using $p_\text{leader}(k=0) = 0$, random initial ego truck position $p_\text{ego}(k=0)$ from a \(- (v_\text{ego}(k=0) \cdot Tg_\text{target} + \text{uniform} (-1.39, 1.39))\), random initial leader speed \(v_\text{leader}(k=0)\) from \(\text{uniform} (8.3, 22.2)\) and \(v_\text{ego}(k=0)\) from \(v_\text{leader}(k=0) + \text{uniform} (-1.39, 1.39)\) distributions, and random desired time gap $Tg$ from a \(\text{uniform} (2, 5)\) distribution. To simplify the setup, we also assume $a_\text{leader}(k) = 0$.
        
\begin{figure}[!htbp]
                \centering
                \includegraphics[width=\linewidth]{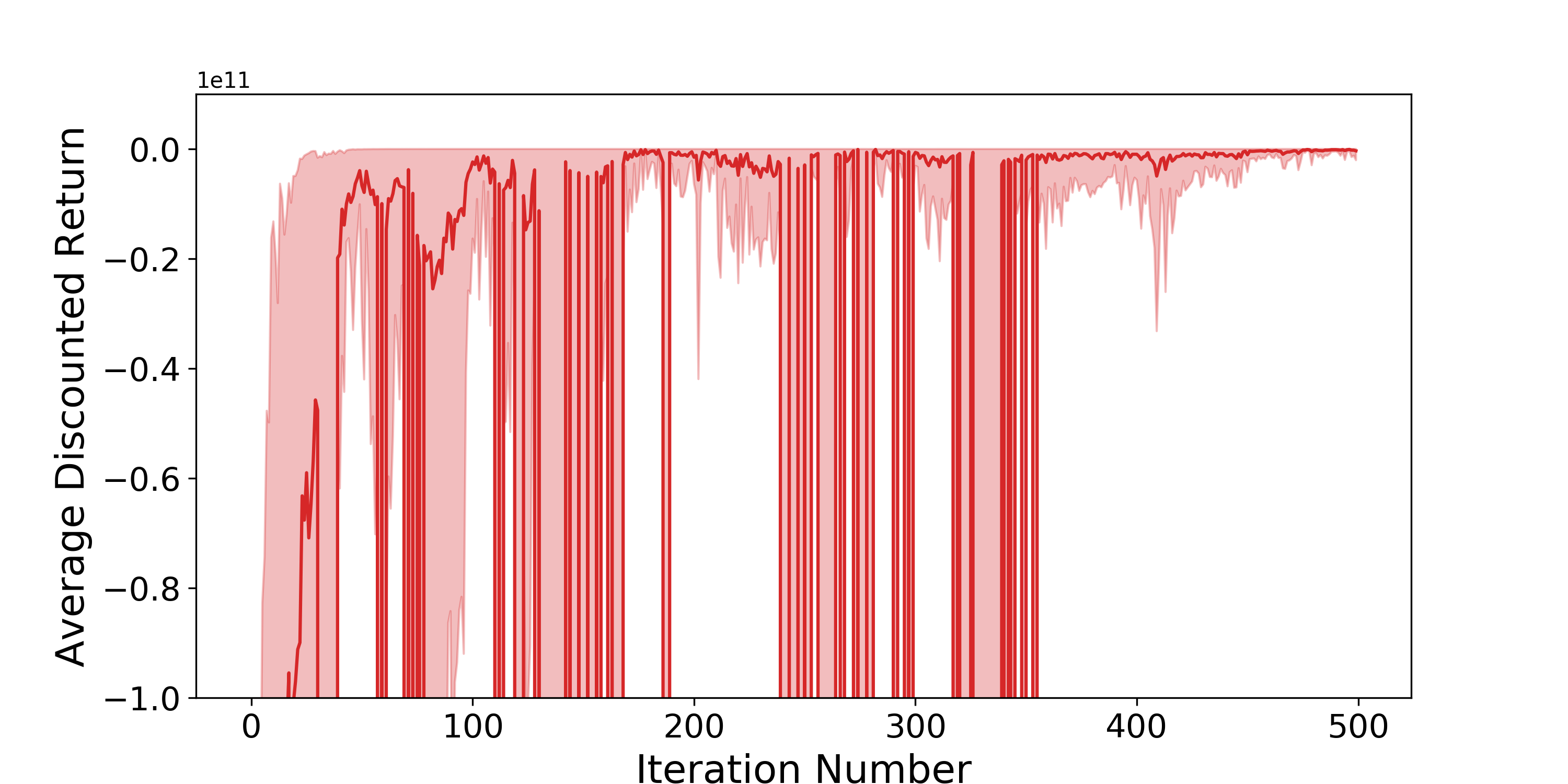}
                \caption{Learning curve---min/max (light red shaded area) and mean (dark red curve) from ten seeds---for deep cruise control policy based on deep model of the TruckSIM truck.}
                \label{fig:trucksim-rl-cacc-learning-curves}
            \end{figure}

            The deep-RL controller was trained on RLLab \cite{duan2016benchmarking} using batch size of 20000, max path length of 800 (sampled at 10Hz) and discount factor of 0.9999. We trained ten policies (ten seeds). The average discounted returns plot is shown in Figure~\ref{fig:trucksim-rl-cacc-learning-curves}. The trained policies shown in the plot converged after 500 iteration. The observed sharp numerical negative infinity return values are caused by crashes between the two trucks inside the training environment as specified by the reward function presented in Section~\ref{sec:deep-rl-cacc}. These crashes happen as the agent of the deep reinforcement learning explores the state-action space, which is implemented here by means of a stochastic agent policy.
            
        \subsubsection{Control validation results}

            \begin{figure*}[!htbp]
                \centering
                \begin{subfigure}[b]{0.49\textwidth}
                    \includegraphics[width=\linewidth, trim={4cm 0 3cm 0},clip]{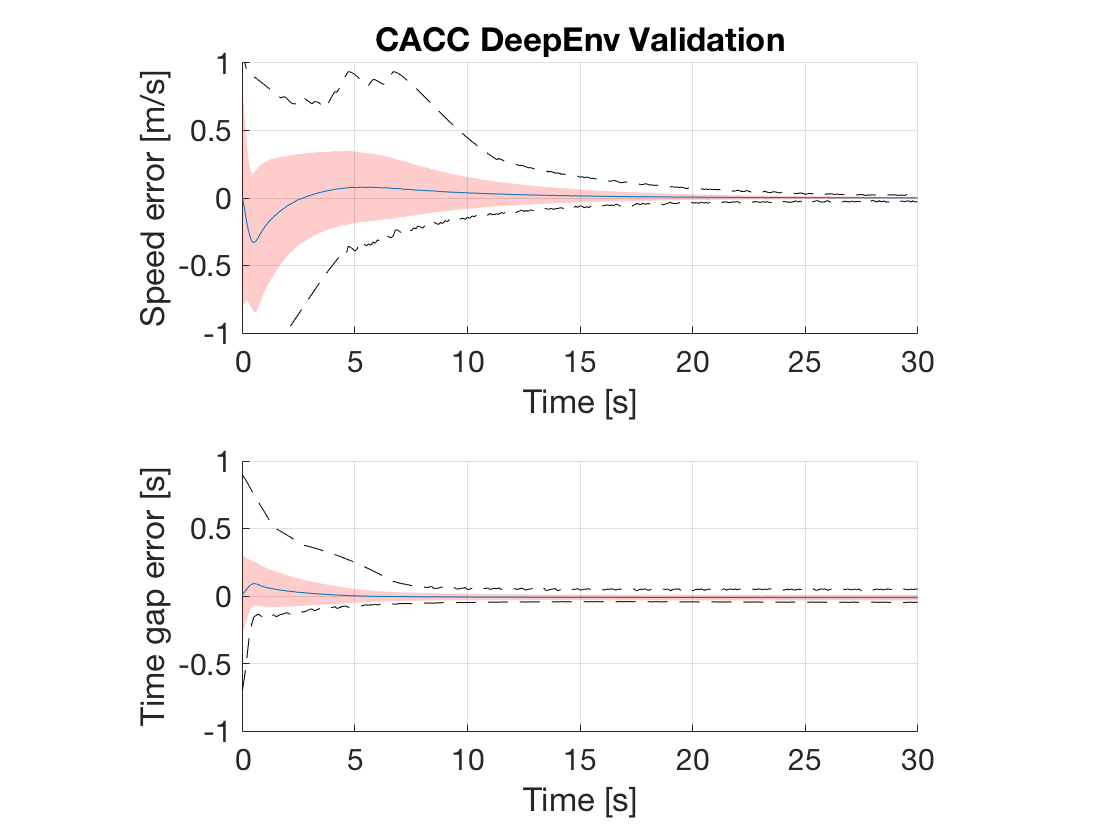}
                \end{subfigure}
                \begin{subfigure}[b]{0.49\textwidth}
                    \includegraphics[width=\linewidth, trim={4cm 0 3cm 0},clip]{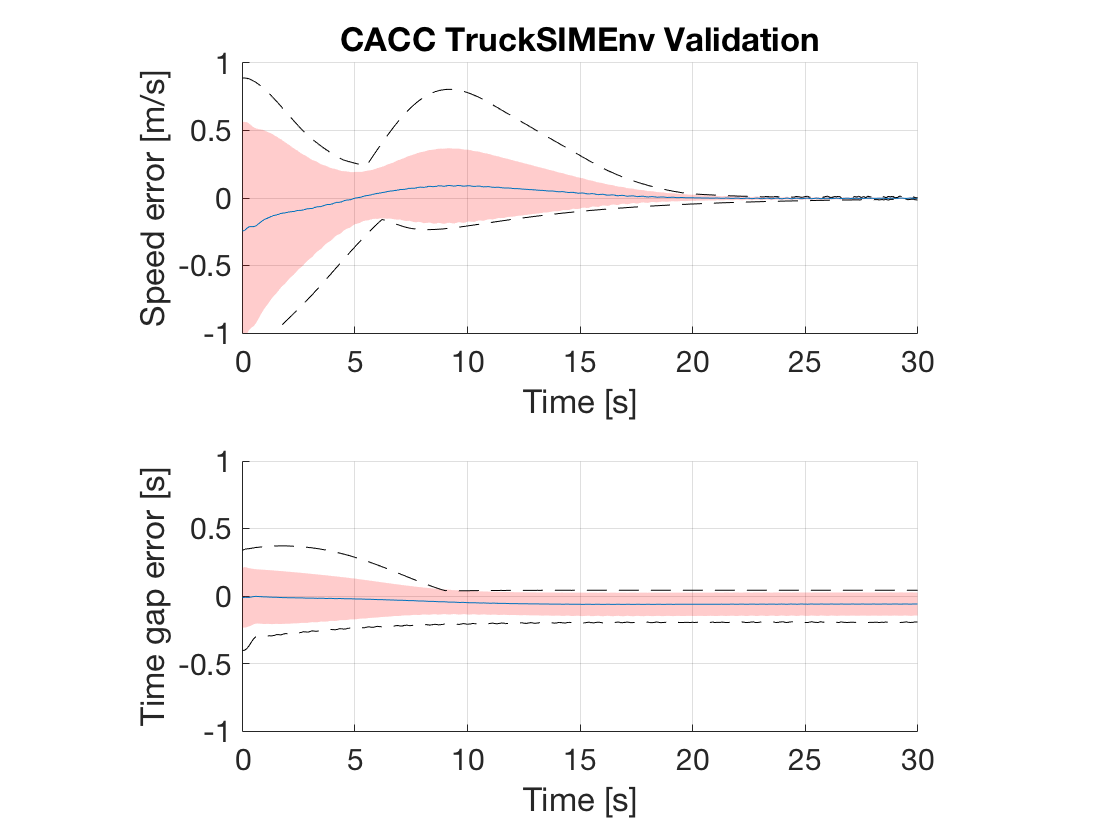}
                \end{subfigure}
                \caption{Control error statistics---min/max (dashed curves), standard deviation (red shaded areas) and mean (blue curves)---for the deep policy evaluated against the deep environment and against TruckSIM environment.} 
                \label{fig:trucksim-truck-rl-cacc-experiment}
            \end{figure*}
            
            This section validates the performance of the deep-RL cooperative adaptive cruise controller against both the deep environment and transfer into TruckSIM as shwon in Figure~\ref{fig:trucksim-truck-rl-cacc-experiment}. DeepEnv-set experiment is a replication of the training setup and consists of 100 rollouts drawn from the same training distributions (environment model and initialization distributions). The same controller is zero-shot transferred to TruckSIM to produce TruckSIM-set consisting of 10 rollouts drawn from the same initialization distributions.
            
            The policy is designed to simultaneously regulate speed and time-gap. In DeepEnv-set, time-gap error converges to steady state error between $\pm0.05s$ within $10s$ from the start time of the experiment, while speed error converges to steady state error between $\pm0.03m/s$ within $25s$ from the start time of the experiment both with mean error of approximately zero. 
            
            TruckSIM-set evaluates the transfer of the same policy to TruckSIM environment with the same random distributions. Observed shift in control performance is caused by shift in truck model distribution due to modal mismatch discussed in the modeling experiments. Time-gap error converges to steady state error between $0.04s$ and $-0.2s$ within $10s$ from the start time of the experiment, while speed error converges to steady state error between $\pm0.02m/s$ within $25s$ from the start time. The time-gap mean error converges to $-0.06s$, while speed mean error converges to approximately zero. The controller exhibits a nonlinear bimodal speed control over/undershoot.
            
            Figure~\ref{fig:trucksim-truck-rl-cacc-experiment} shows preliminary learning results for deep-RL cooperative adaptive cruise controller and shows preliminary evidence to expect marginal shifts in error statistics when transferring the policy from the deep-truck environment to the ``real'' environment (here conducted using a simulated truck).
             
            
    
    \subsection{Deep modeling of full-size trucks (field experiments)}
        This section presents field experimental results for the model described in this article. The section documents experiments conducted using two differently configured real-physical heavy duty trucks. These same two trucks were modeled using two different physics-based power-train models in \cite{XYLu2005HDVModelandLongControl} and \cite{lu2017integrated} used to develop high precision control systems within each respective article.
        
        \subsubsection{Configuration one: Freightliner}

            In this experiment, the truck is not equipped with any sensors relevant to the driving environment (e.g. road grade) and thus we substitute $w(k)$ with the empty set. The controllable input to the truck is given by $u(k) = [ E_\text{cmd}(k), B_\text{cmd}(k)], $ where $E_\text{cmd}(k)$ is requested percentage engine torque in [$0-100\%$] and $B_\text{cmd}(k)$ is service brake pedal position [$0-100\%$].
            
            
            The output (truck response) vector is given by $y(k) = [a(k), v(k), F_\text{rate}(k)],$ where $a(k)$ is longitudinal acceleration in [$m/s^2$], $v(k)$ is longitudinal speed in [$m/s$], and $F_\text{rate}(k)$ is fuel rate in [$cm^3/s$].
        
            Experiments for this truck has been carried out at a nearly flat test track with straight roads the longest of which is around 300 meters long at the Richmond Field Station at California. The truck was driven for about 16 minutes to collect primarily slow speed dataset covering from zero to $18m/s$. The dataset was split into 85 percent for training and 15 percent to test modeling performance on an unseen dataset.
            
                
        \subsubsection{Configuration two: Volvo}
    
            The truck is equipped with a road grade sensor where $w(k) = \theta_\text{rdg}(k) [\%]$. The controllable input to the truck is given by $u(k) = [E_\text{cmd}(k), B_\text{cmd}(k)],$ where $E_\text{cmd}(k)$ is requested percentage engine torque in [$0-100\%$] and $B_\text{cmd}(k)$ is service brake command [$m/s^2$].
            
            Actuation and accessible signal measurement of the brake system in this truck is asymmetric. The service brake system in this truck is not directly actuatable (and signals not interceptable); instead, desired deceleration is processed through Volvo propriety systems. After collecting the data, we substitute brake commands with observed deceleration gated by brake pedal gating switch signal. 
            
            The output (truck response) vector is given by $y(k) = [a(k), v(k), F_\text{rate}(k)],$ where $a(k)$ is longitudinal acceleration in [$m/s^2$], $v(k)$ is longitudinal speed in [$m/s$], and $F_\text{rate}(k)$ is fuel rate in [$cm^3/s$].
            
            This truck was primarily driven over non-flat open freeways. The truck was driven for about 24 minutes to collect primarily freeway speed driving dataset covering speeds from $20m/s$ to $30m/s$. The dataset was split into 85 percent for training and 15 percent to test modeling performance on an unseen dataset.
            
                
        \subsubsection{Results and model validation}
\begin{figure*}[!htbp]
                \centering
                \includegraphics[width=\linewidth, trim={4cm 0 3cm 0},clip]{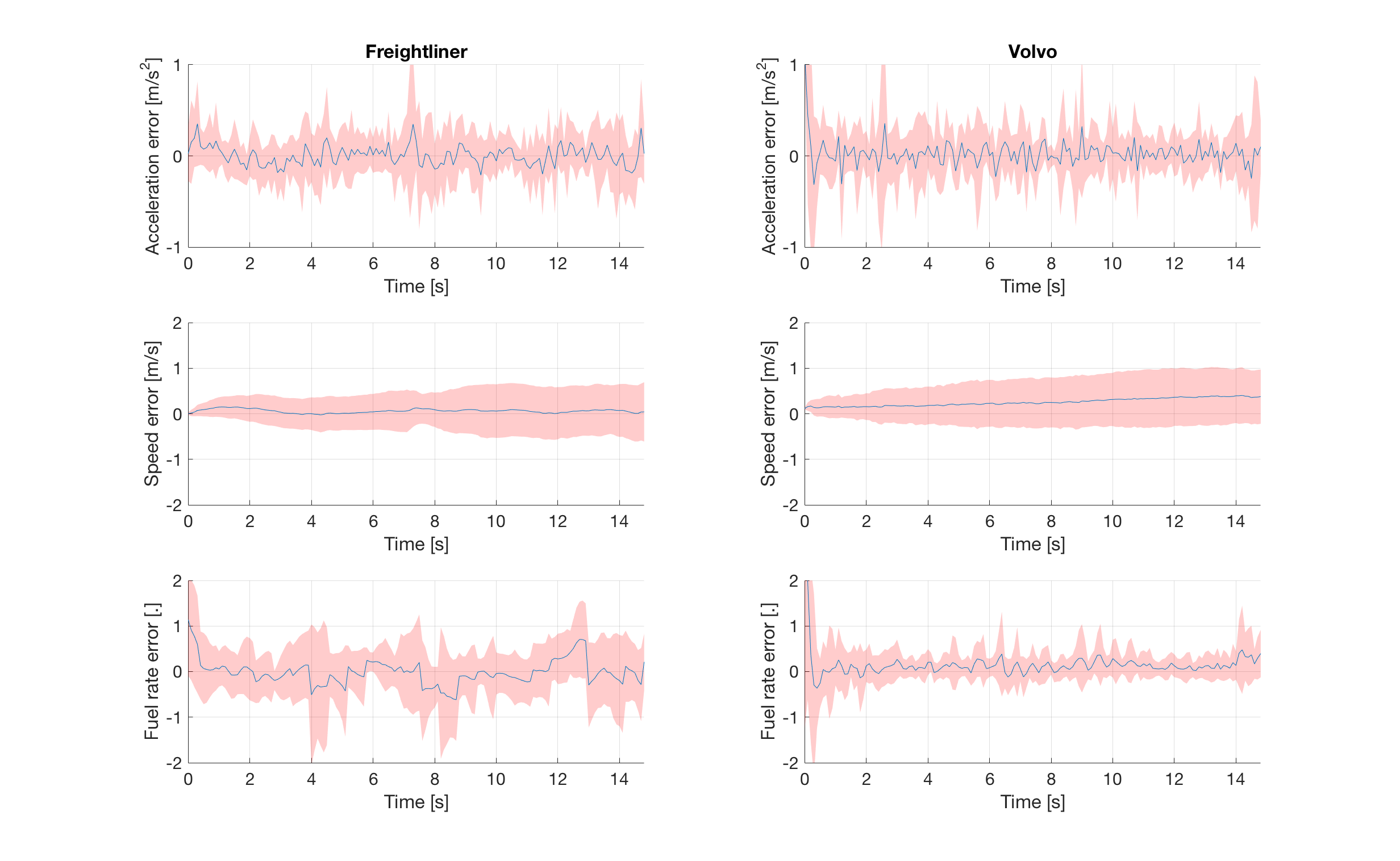}
                \caption{Model error statistics---standard deviation (red shaded areas) and mean (blue curves) from 10 trials---for deep modeling of the Freightliner and the Volvo trucks.}
                \label{fig:field-experiments-error-stats}
            \end{figure*}

            We validate modeling performance against an unseen ground truth dataset from each truck configuration as shown in Figure~\ref{fig:field-experiments-error-stats}. In this figure, mean and standard deviation for acceleration, speed, and fuel rate modeling errors are charted as a function of model simulation time. The statistics were produced from an ensemble of ten timeseries simulations. Each simulation is fresh initialized at time zero, and simulated using knowledge of inputs and the uncontrollable conditions only. Error statistics are generated as $\text{ErrorStatistic}(k)= \text{Statistic}_m\left(\hat{y}_{m}(k| \Phi) - y_{m}(k)\right)$ where $m$ is trial number, and $\hat{y}$ follow the deployment form Equation~\eqref{eq:model-deployment-form}.
            
            In this figure, acceleration error is bounded between $\pm0.5m/s^2$. For the Freightliner, speed error remained bounded between $\pm0.5m/s$ mean of speed error $0.12m/s$ after the initial transient. For the Volvo, speed error remained between $-0.5m/s$ and $1m/s$ with a significant error bias approaching $0.5m/s$ during the 15 seconds of simulation. Fuel rate modeling error is bounded between $\pm1$ once the initial transient decays. We speculate that model performance degradation for the Volvo truck is influenced by insufficient data to model truck dynamics over graded roads.
            
        
    \subsection{Deep-RL control of full-size trucks (field experiments)}
        This section presents control experiments for the deep-RL CACC system presented earlier. Due lab access limitations during the CoVID-19 pandemic, the system was operated as a two vehicle ACC (using radar instead of direct communications) system on non-flat open freeways. The leader is a passenger car and the follower is the Volvo truck presented earlier.
        
\begin{figure*}[!htbp]
            \centering
            \begin{subfigure}[b]{0.49\textwidth}
                \includegraphics[width=\linewidth, trim={4cm 0 3cm 0},clip]{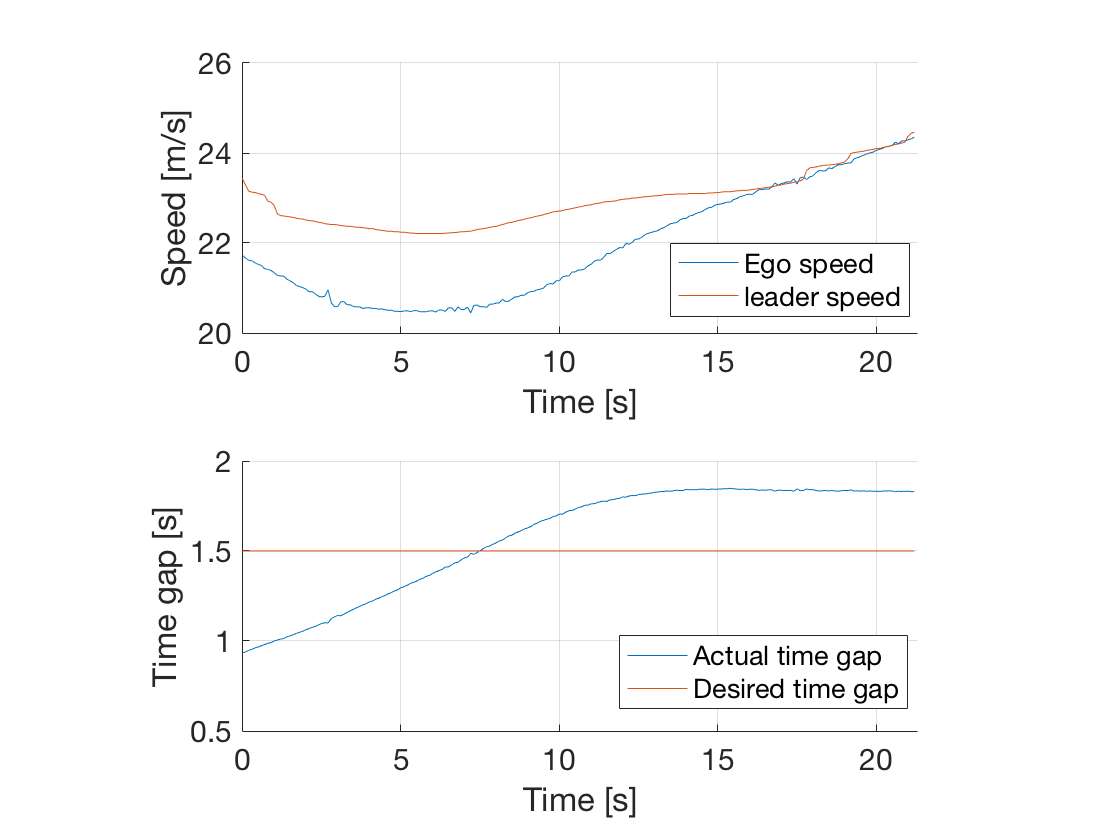}
            \end{subfigure}
            \begin{subfigure}[b]{0.49\textwidth}
                \includegraphics[width=\linewidth, trim={4cm 0 3cm 0},clip]{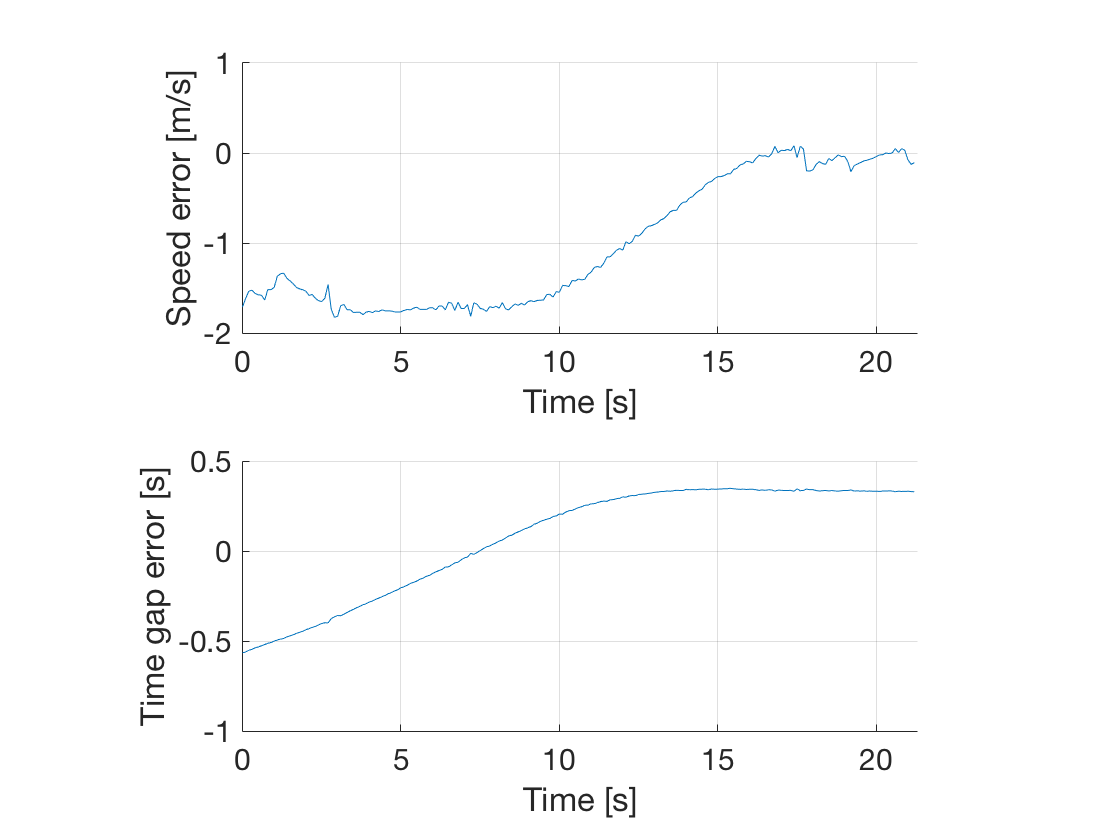}
            \end{subfigure}
            \caption{Tracking speed, time-gap, and control error for the Volvo deep CACC policy evaluated against the real environment---gap closing maneuver.} 
            \label{fig:volvo_rl_acc_gap_closing}
        \end{figure*}

        Figure~\ref{fig:volvo_rl_acc_gap_closing} shows gap closing regulation performance where the leader drove at nearly constant speed with initial speed error of $2m/s$, initial time gap error of $1s$, and a desired time-gap setting of $1.5s$. The gap was closed within 15 seconds and to within error bound of $\pm0.2m/s$ and $0.35s$. Leader conducted a quick successive changes of speed towards the end of experiment causing the observed speed ripple after time $17s$.
        
\begin{figure*}[!htbp]
            \centering
            \begin{subfigure}[b]{0.49\textwidth}
                \includegraphics[width=\linewidth, trim={4cm 0 3cm 0},clip]{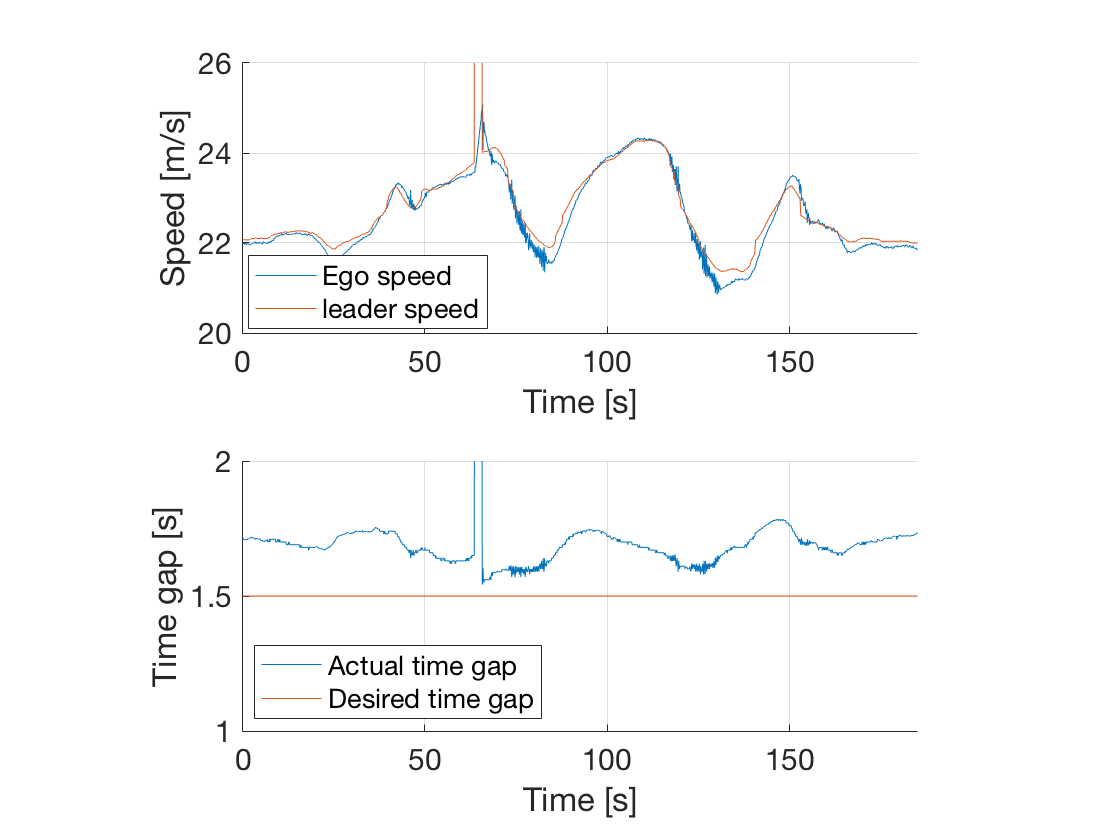}
            \end{subfigure}
            \begin{subfigure}[b]{0.49\textwidth}
                \includegraphics[width=\linewidth, trim={4cm 0 3cm 0},clip]{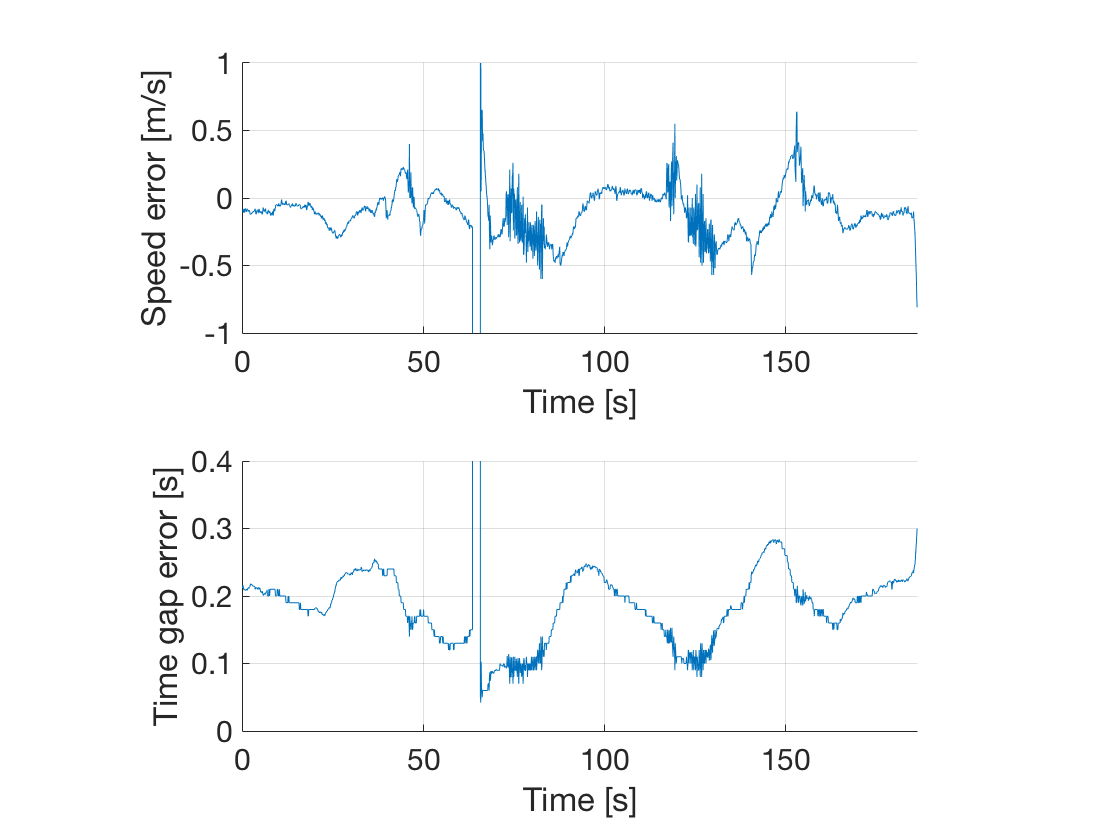}
            \end{subfigure}
            \caption{Tracking speed, time-gap, and control error for the Volvo deep CACC policy evaluated against the real environment---leader following maneuver.} 
            \label{fig:volvo_rl_acc_long_run}
        \end{figure*}

        Figure~\ref{fig:volvo_rl_acc_long_run} shows tracking performance over an arbitrary driving cycle conducted by the leader vehicle with a desired time-gap setting of $1.5s$. Speed error was regulated to within $\pm0.5m/s$ and time gap was regulated to between $0.05m/s$ and $0.3m/s$. A lane change maneuver was conducted at time $63s$ causing a momentary misalignment between ego vehicle's sensor line-of-sight with the leader. Speed and distance measurements of a farther vehicle down stream was detected causing the observed discontinuity.
        
        
        
\section{Conclusion}
    Detailed study of each heavy duty truck in some pool of trucks has historically been required to develop and fit precise analytical models and controls. This article discusses the application of deep learning and deep reinforcement learning as an approach to simplify the process and abstract detailed vehicle underlying mechanics with a potential for improving modeling and control precision. A brief experimental evaluation is presented as a walk through the process and as preliminary performance validation. 

    The deep models and deep-RL controls presented in this article successfully (1) infers relevant latent and state variables (such as gearbox), (2) performs dynamic state estimation (such as selected gear and brake cylinder pressure at \(t=0\)) and tracking (latent state variable values for \(t>0\)), and (3) successfully performs system identification and parameter estimation (such as the aerodynamic drag effect and its coefficient).
    
    This article focuses on outlining the process of applying deep learning and deep reinforcement learning for modeling and control of heavy duty trucks. More extensive experimentation and comparison with established classical approaches is still required for validation and performance evaluation. Furthermore, the process presented here still requires full replication for each target truck, and each truck combination (multi-truck environments). Further investigation is still required to introduce transfer learning of longitudinal dynamics across mechanical configurations. Data sampling efficiency and utilization of existing first-principle models could also be investigated to improve the process presented here.
    
\section*{Acknowledgements}
    The authors would also like to acknowledge John Spring and David Nelson for their technical conversations about automation software and hardware for heavy duty trucks and their support in the field. This research work was supported in part by King Abdulaziz City for Science and Technology (KACST).

\bibliographystyle{apalike}
\bibliography{references}

\end{document}